\documentclass[%
twocolumn,
superscriptaddress,
 amsmath,amssymb,
 prx,
showkeys,
showpacs,
longbibliography
]{revtex4-2}
\usepackage{graphicx}
\usepackage{amsmath, amssymb, amsfonts, bm}
\usepackage{latexsym}
\usepackage{hyperref}
\hypersetup{
    unicode=false,          % non-Latin characters in AcrobatÕs bookmarks
    pdftoolbar=false,        % show AcrobatÕs toolbar?
    pdfmenubar=true,        % show AcrobatÕs menu?
    pdffitwindow=false,     % window fit to page when opened
    pdfstartview={FitH},    % fits the width of the page to the window
    pdftitle={},    % title
    pdfauthor={},     % author
    pdfsubject={},   % subject of the document
    pdfcreator={},   % creator of the document
    pdfproducer={}, % producer of the document
    pdfnewwindow=true,      % links in new window
    colorlinks=true,       % false: boxed links; true: colored links
    linkcolor=blue,          % color of internal links (change box color with linkbordercolor)
    citecolor=blue,        % color of links to bibliography
    filecolor=magenta,      % color of file links
    urlcolor=blue           % color of external links
}
\usepackage{mathdots}
\usepackage[protrusion=true,expansion=true]{microtype}
\usepackage{suffix}
\usepackage{makecell}
\usepackage{float}
\usepackage{slashed}
\usepackage{siunitx}
\usepackage{mathrsfs}

\newcommand{\ii}{\mathrm{i}}
\newcommand{\D}[1]{\,\text{d}#1\,}

\newcommand{\bra}[1]{\mathinner{\langle{#1}|}}
\newcommand{\ket}[1]{\mathinner{|{#1}\rangle}}

\newcommand{\braket}[2]{\langle #1 | #2 \rangle}

\newcommand{\matel}[3]{\langle #1 | #2 | #3\rangle}

\newcommand{\fixme}[1]%
   {\begingroup{\color{blue}[NOTE: \textit{#1}]}\endgroup}

\newcommand{\nop}{\hat n}
\newcommand{\aop}{\hat a}

\newcommand{\hop}{\hat H}
\newcommand{\lop}{\hat L}
\newcommand{\rop}{\hat R}
\newcommand{\sop}{\hat S}
\newcommand{\dens}{\hat\rho}

\newcommand{\uop}{\hat U}
\newcommand{\vop}{\hat V}
\newcommand{\dop}{\hat D}

\newcommand{\qub}{\text{q}}
\newcommand{\cent}{\text{c}}
\newcommand{\tc}{\text{c}}

\newcommand{\sgn}{\textrm{sgn}}

\DeclareMathOperator{\tr}{Tr}
\DeclareMathOperator{\erf}{erf}

\let\vec\oldvec
\newcommand{\vec}{\mathbf}
\newcommand{\mat}{\textsf}

\usepackage{scalerel,stackengine,amsmath}
\newcommand\equalhat{\mathrel{\stackon[1.5pt]{=}{\stretchto{%
    \scalerel*[\widthof{=}]{\wedge}{\rule{1ex}{3ex}}}{0.5ex}}}}

\begin{document}
\title{Two-qubit-gate operation in a high-connectivity transmon lattice utilizing a tunable coupling to a shared mode}
\author{Tuure Orell}
\affiliation{IQM Quantum Computers, Oulu 90590, Finland}

\author{Hao Hsu}
\affiliation{IQM Quantum Computers, Munich 80992, Germany}

\author{Joona Andersson}
\affiliation{IQM Quantum Computers, Espoo 02150, Finland}

\author{Jani Tuorila}
\affiliation{IQM Quantum Computers, Oulu 90590, Finland}

\author{Frank Deppe}
\affiliation{IQM Quantum Computers, Munich 80992, Germany}

\author{Hsiang-Sheng Ku}
\affiliation{IQM Quantum Computers, Munich 80992, Germany}
\date{October 2025}

\begin{abstract}
    Increasing connectivity and decreasing qubit-state delocalization without compromising the speed and accuracy of elementary gate operations are topical challenges in the development of large-scale superconducting quantum computers. In this theoretical work, we study a special honeycomb qubit lattice where each qubit inside a unit cell is coupled to every other one via two dedicated tunable couplers and a common central element. This results in an effective multi-mode interaction enabling tunable, on-demand, all-to-all connectivity between each qubit pair within the unit cell. 
    We provide a thorough analysis of the unit cell, including a proposal for a novel and efficient conditional-Z gate scheme which takes advantage of the effective multi-mode coupling. We develop an experimentally viable pulse protocol for a single-step gate implementation which considerably improves the gate speed compared to the previous two-qubit-gate realizations suggested for architectures utilizing a center mode. 
    We also show numerical results on how the presence of spectator qubits affects the average two-qubit-gate fidelity, and analyse how the multi-mode coupling structure mitigates the delocalization-induced crosstalk during simultaneous single-qubit gates within the unit cell. We also provide analytical estimates for the errors caused by relaxation and dephasing during a two-qubit-gate operation, including noise terms for the multi-mode coupling structure.
    Our multi-mode coupling architecture results in a good balance between increased connectivity and available parallelism, especially when several interacting unit cells form a quantum processing unit. We anticipate that the obtained results pave the way towards high-connectivity quantum processors with efficient and low-overhead quantum algorithms. 
\end{abstract}

\date{\today}
\maketitle

\section{Introduction}

Conventional superconducting quantum-processor architectures are built upon relatively low-connectivity planar grids such as the square lattice~\cite{arute2019, Wu2021, Abdurakhimov2024,Acharya2025} and the heavy-hexagonal lattice~\cite{Hertzberg2021}. In general, lower connectivity results in longer algorithm execution times since more elementary gate operations are needed to realize the desired quantum circuit~\cite{Holmes2020}. Another emerging limitation to quantum-processor operation in the superconducting framework is the quantum (delocalization, hybridization) crosstalk~\cite{Zhao2022b} arising from the fact that the control signals couple to local degrees of freedom whereas the computational states are delocalized eigenstates of the whole processor~\cite{Sung2021}. Delocalization causes errors in particular for single-qubit gates, and can become a dominating source of quantum-processor infidelity especially if the other error channels, such as leakage and classical crosstalk, are mitigated with novel pulse-shaping techniques~\cite{Hyyppa2024}.

Delocalization of the computational states can potentially be mitigated with tunable couplers~\cite{Yan2018,  Li2020, Heunisch2023a}, and can further be suppressed by adding extra modes in the coupling elements~\cite{McKay2015, Jiang2025}. Typical tunable couplers have one mode~\cite{Yan2018}, and there has been several high-fidelity demonstrations of single- and two-qubit gate operations in superconducting two-qubit test devices~\cite{Marxer2023,marxer2025,Hyyppa2024} and also on larger low-connectivity quantum processors with tunable couplers~\cite{arute2019,Acharya2025, AbuGhanem2025}. However, single-mode couplers cannot suppress delocalization completely~\cite{Heunisch2023a} and, thus, the computational states of such devices are still susceptible to delocalization-induced errors. Such errors may become detrimental in algorithms during which the neighbouring spectating qubits are excited, for example when highly-entangled multi-qubit states are needed such as in Shor's algorithm~\cite{Shor1997}.

In large-scale quantum computers with single-mode-coupler architecture, a practical design choice is to use a square-grid topology, which limits the connectivity to four nearest-neighbour qubits in the bulk of the processor. However, many algorithms require long-range interactions between qubits, which are then implemented as a sequence of nearest-neighbour gates. Consequently, errors from the sequential gates accumulate, resulting in a reduced final gate fidelity compared to that of the individual gates. This problem has been mitigated using multi-qubit couplers~\cite{Song2017, Song2019, Hazra2021, Huber2025}, where several qubits are coupled directly to a shared mode, for example a resonator, without additional tunable couplers in between. Despite the increase in connectivity, such implementations are even more vulnerable to delocalization errors than those based on two-qubit couplers, due to the increased number of qubits per coupling mode.

There has also been proposals for two-mode couplers~\cite{mundada2019,Moskalenko2021, Goto2022,Wu2024a, Li2024,Jiang2025} which can potentially be used to mitigate the delocalization and the related errors within the superconducting quantum computing framework. However, using two-mode couplers in high-connectivity architectures appears challenging, e.g. because of the susceptibility to spectator errors when the number of connected qubits is scaled up. On the other hand, increasing the number of coupling modes indefinitely appears also harmful since each additional mode increases the number of leakage, dissipation and decoherence channels in the system. We thus conclude that a promising choice is to have~$N+1$ coupling modes to realize an all-to-all connectivity between~$N$ transmon qubits.

In this paper, we study the multi-mode coupler framework by building upon our previous work~\cite{Renger2025} in which we studied a setup of $N=6$ transmon qubits that are coupled through their own dedicated tunable couplers to a shared resonator mode. The shared center mode mediates pair-wise qubit interactions and acts as an additional filter for the unwanted delocalization-induced crosstalk and leakage to spectator qubits. 
In Ref.~\onlinecite{Renger2025}, an effective two-qubit conditional-Z (CZ) gate was created using the so-called MOVE-CZ-MOVE protocol. However, since the constituent operations are done sequentially (‘MOVE’ + ‘CZ’ + ’MOVE’), the total gate time is unavoidably long compared to the conventional diabatic CZ gate in square lattice topology~\cite{Marxer2023,marxer2025}. 

Here, we introduce a novel pulse scheme for the realization of a CZ gate mediated by a multi-mode coupling between two qubits in a unit cell of a honeycomb lattice. Our approach relies on simultaneous excitation swaps in single- and two-excitation manifolds of the setup consisting of the two qubits, two couplers and the center mode, thus allowing efficient single-step calibration with a duration in principle equal to two MOVE operations. Consequently, the gate operation time is considerably reduced compared to the MOVE-CZ-MOVE protocol.

In the honeycomb lattice, each qubit is coupled to three center modes, and thus can be coupled to the 12 nearest neighbours, as shown in Fig.~\ref{fig:fig1}. Inside the unit cell only one CZ gate can be performed at a time. But, since in a lattice formed by unit cells there are on average two qubits per a center mode, it is possible to perform CZ gates in parallel such that all the qubits within the bulk of the lattice can be operated simultaneously. Such optimized connectivity has applications in quantum error correction, more specifically in color codes, quantum low-density parity-check (qLDPC) codes, and tile codes~\cite{Vigneau2025, Steffan2025, breuckmann2025, yang2025}. Furthermore, by implementing the central mode with a coplanar waveguide resonator, one could facilitate a long-range coupler required for example by the qLDPC codes~\cite{choe2026}.

We provide an exhaustive analytical and numerical analysis of the novel gate scheme, including optimization of the pulse parameters and analysis of the error mechanisms. We describe in detail the theoretical and numerical methods required to obtain the presented results. We also discuss the potential challenges with this approach and provide mitigation solutions when available. In order to keep things as simple as possible, we concentrate on pulses with simple analytical forms. Using more sophisticated pulse shapes, such as Slepian~\cite{Martinis2014} for the two qubit gates, and DRAG and its extensions~\cite{Motzoi2009, Chow2010, Hyyppa2024} for single-qubit gates, could be used to mitigate different errors even further, resulting into faster gates.

The paper is organized as follows. In Sec.~\ref{sec:system} we define the system Hamiltonian. In Sec.~\ref{sec:gate} we present the basic operation principle of our high-connectivity CZ gate and demonstrate its performance in a minimal system consisting of two qubits and the coupling element. In Sec.~\ref{sec:spectator_hybrid} we discuss how spectating qubits and hybridization affect the accuracy of the gate. Section~\ref{sec:disspation} briefly discusses the effects of relaxation and dephasing, and the conclusive remarks are given in Sec.~\ref{sec:conclusions}.

\section{System and Hamiltonian}\label{sec:system}

\begin{figure*}
    \centering
    \includegraphics[width=0.9\linewidth]{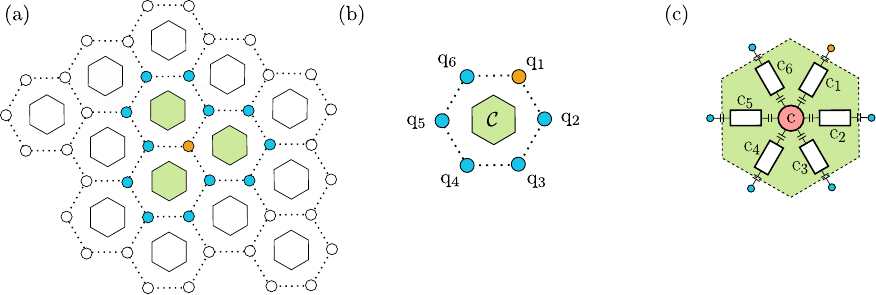}
    \caption{Schematic of a quantum-processing unit with the honeycomb topology utilizing multi-mode couplers.~(a)~Schematic of the honeycomb lattice with multi-mode couplers. Ideally, there are no direct qubit-qubit couplings along the unit cell boundaries (dashed lines), and each qubit is capacitively coupled only to the three nearest multi-mode couplers. We have highlighted the nearest-neighbour qubits (blue) and multi-mode couplers (green) for qubit~$\qub_1$ (orange).~(b)~Schematic of the unit cell, where $\mathcal{C}$ denotes the multi-mode coupler.~(c)~Schematic of the multi-mode coupler, with $\tc_i$ representing the $i$th tunable coupler.}
    \label{fig:fig1}
\end{figure*}
We consider here one unit cell of a quantum processing unit where the qubits are arranged in the honeycomb lattice topology, as shown in Fig.~\ref{fig:fig1}. The qubits in the unit cell of the lattice are coupled with a tunable multi-mode coupler which allows on-demand coupling between any qubit pair in the unit cell while keeping the couplings to spectating qubits suppressed. Here, we consider in more detail a specific superconducting realization of the setup in which each lattice site hosts a transmon qubit. The multi-mode coupling between the transmons is created by a central transmon qubit coupled to the $N=6$ peripheral qubits through dedicated tunable transmon couplers~\cite{Yan2018, Marxer2023} which allow the control of the coupling strengths between the peripheral qubits and the central transmon. Each transmon in the system includes a SQUID loop which allows tuning of the transmon frequency with external magnetic flux threading the loop. We emphasize here that the results and conclusions made in this paper apply also to a case in which the central anharmonic mode is replaced by a linear resonator~\cite{Renger2025}. 

We treat the transmons as weakly anharmonic Duffing oscillators and, consequently, the system is described by the Hamiltonian
\begin{align}
    \hop / \hbar &= \sum_{\ell}\left[\omega_\ell \nop + \frac{\alpha_\ell}{2}\nop_\ell(\nop_\ell-1)\right]\nonumber
    \\
    &- \sum_{\ell\neq\ell'}g_{\ell\ell'}\left(\aop_\ell^\dag-\aop_\ell\right)\left(\aop_{\ell'}^\dag-\aop_{\ell'}\right),\label{eq:ham1}
\end{align}
where~$\aop_\ell$ and~$\aop_\ell^\dag$ are the annihilation and creation operators of the (local) mode~$\ell$,~$\nop_\ell = \aop_\ell^\dag\aop_\ell$ is the corresponding number operator, and~$\omega_\ell$ and~$\alpha_\ell$ are the angular frequencies and anharmonicities (for transmons~$\alpha_\ell$ is negative). The coupling strength~$g_{\ell\ell'}$ between the modes~$\ell$ and~$\ell'$ depends on the mode frequencies as~$g_{\ell\ell'}\equiv\beta_{\ell\ell'}\sqrt{\omega_\ell\omega_{\ell'}}$, where~$\beta_{\ell\ell'}$ is a dimensionless constant the value of which is determined by the capacitance matrix of the system~\cite{Yan2018}. 

The qubits are coupled to their respective tunable couplers and to the shared center mode, but not to each other. Thus, labeling the qubit modes as~$\qub_j$, tunable coupler modes as~$\tc_j$ and the center mode as~$\cent$, the only non-zero couplings in the Hamiltonian~\eqref{eq:ham1} are~$g_{\cent\qub_j}$,~$g_{\qub_j\tc_j}$ and~$g_{\cent\tc_j}$, with~$j$ running from 1 to~$N$, so that in total we have~$2N+1$ modes in the system. For example, in order to conduct a two-qubit gate operation between qubits q$_1$ and q$_2$ in the unit cell (see Fig.~\ref{fig:fig1}), we tune the corresponding local coupler frequencies $\omega_{\tc_1}$ and $\omega_{\tc_2}$ to the operation values for which the coupling between the qubits q$_1$ and q$_2$ and the center mode becomes strong, thus creating an effective three-mode coupling between the qubits. 

In reality, parasitic capacitive couplings exist between the modes — for instance, direct couplings between the tunable couplers or between the peripheral qubits — which we neglect here for simplicity. These couplings can increase population leakage to spectator qubits during two-qubit gates, particularly when they are near resonance with the qubits involved in the gate. It is therefore usually beneficial to introduce some disorder in the qubit idling frequencies, which reduces the unwanted population swapping caused by these parasitic interactions.

We calculate results for a varying number of qubits between~$N=2$ to~$N=6$. This is, for example, to better identify the effects arising from the spectating qubits. The numerical results are computed using the parameters in~Tab.~\ref{tab:parameters}, unless stated otherwise. The numerical values of the parameters chosen here are experimentally relevant and viable, based partly on measured values (qubit frequencies and anharmonicities) and partly on design values (couplings and coupler anharmonicities) of a device currently under development~\cite{iqm}. Similar values are used also in the literature~\cite{Marxer2023, marxer2025, Renger2025, sete2021}. Importantly, the gate scheme studied here performs well over a wide range of these parameters, as shown in App.~\ref{app:ham_par}, so the qualitative conclusions of this work do not depend sensitively on the specific device parameters chosen here.

\subsection{Computational basis}

Hamiltonian in Eq.~\eqref{eq:ham1} describes a system of coupled anharmonic modes. We observe that the cardinality of the Hilbert space is much larger (infinite) than is required for universal quantum information processing typically defined in terms of two-level systems. Moreover, the couplings occur between the eigenstates of the local number operators~$\hat n_\ell$, thus creating delocalization in the eigenbasis of the coupled system. Ideally the computational basis of a quantum computer is defined as the joint eigenbasis of Pauli-Z operators of all qubits. However in our case, the different cardinality together with the delocalization generated by the couplings, makes the definition of the computational basis somewhat subtle issue. We follow the practice of choosing the computational basis such that it is appropriate for practical computation. This means that one has to be able to conduct the basic operations between the computational states fast and with high accuracy. Such operations include those listed in the DiVincenzo criteria for quantum computation~\cite{divincenzo2000}, i.e. the capability of initializing the system to some computational state, conducting universal gate operations, and ability to read out the state in the computational basis.   

The definition of the computational basis of a general interacting anharmonic oscillator system is not unique, but depends for example to which operators the available control signals couple. Here, we concentrate on superconducting transmon implementations of the honeycomb lattice with the multi-mode coupling mediated by the center element. Using the conventional design, the microwave signals controlling single-qubit-gate operations and the readout resonators couple to the local charge operators of the transmons defined as~$\hat q_\ell = \ii q_\ell^{\rm zp}(\hat a_\ell^\dag-\hat a_\ell)$ where~$q_\ell^{\rm zp}$ are the zero-point fluctuations of the local charges. Additionally, one often has the capability to control the local frequencies of the qubits and the couplers using local external fluxes which couple to the system through the local number operators~$\hat n_\ell$. Conventional readout techniques probe the qubit states in the eigenbasis of the coupled system~\cite{Sung2021}, despite the fact that the readout resonators are coupled to the local charge operators of the transmons. Moreover, the conditional phase collected during CZ gates is typically defined between eigenstates of the coupled system. Therefore, we follow the convention~\cite{Galiautdinov2012, Chu2021,Sung2021} that the computational basis states are eigenstates of the Hamiltonian defined in Eq.~\eqref{eq:ham1} for a chosen set of frequencies~$\omega_\ell$. For simplicity we neglect the readout resonators in this study.

\subsection{Delocalization, longitudinal coupling and idling configuration}\label{sec:idling}

Even though the computational basis states are eigenstates of the Hamiltonian~\eqref{eq:ham1}, there can still be complications to idling or gate fidelities of the qubits caused by the couplings between localized states. This arises from the fact that the eigenfrequencies are coupling-dressed local frequencies, and the eigenstates are superpositions of several local number operator eigenstates, a phenomenon which is often referred to as hybridization or delocalization. Dressing in the eigenfrequencies can lead to unwanted conditional-phase errors during idling and, since the single-qubit-gate drive signals couple to the local transmon operators, hybridization can induce quantum cross-talk~\cite{Zhao2022b} that results in leakage during simultaneous single-qubit gates, or during single-qubit gates if spectator qubits are excited. Consequently, one has to pay detailed attention to finding a proper idling configuration for the qubit and coupler frequencies such that the conditional phase rates and delocalization are as minimal as possible. 

We here present the calibration of the idling configuration in a simplified system and then generalize the findings to our honeycomb unit cell. Let us consider the tunable coupling between the center mode~$\cent$ and one of the qubits, say~$\qub_1$, in the unit cell. These are connected via the tunable coupler~$\tc_1$. We expect that the dressing and delocalization are minimized in the dispersive regime, i.e. if~$g_{\cent \tc_1}\ll |\omega_\cent-\omega_{\tc_1}|$,~$g_{\qub_1 \tc_1}\ll |\omega_{\qub_1}-\omega_{\tc_1}|$, and~$g_{\cent \qub_1}\ll |\omega_\cent-\omega_{\qub_1}|$. 
Similar to Ref.~\onlinecite{Yan2018}, we eliminate the coupler with the Schrieffer--Wolff transformation expanded to the second order in the coupling strengths~$g_{\cent \tc_1}$ and~$g_{\qub_1\tc_1}$, see App.~\ref{sec:SF} for details. Consequently, the effective XY-coupling between the qubit~$\qub_1$ and the center mode~$\cent$ is given as
\begin{equation}
    \widetilde g_{\cent \qub_1} = g_{\cent \qub_1} + \frac{g_{\qub_1 \tc_1}g_{\cent \tc_1}}{2}\left[\frac{1}{\Delta_{\qub_1\tc_1}}+\frac{1}{\Delta_{\cent\tc_1}}-\frac{1}{\Sigma_{\qub_1\tc_1}}-\frac{1}{\Sigma_{\cent\tc_1}}\right],
\end{equation}
where~$\Delta_{\ell \ell'}=\omega_\ell-\omega_{\ell'}$ and~$\Sigma_{\ell \ell'}=\omega_\ell+\omega_{\ell'}$. We note that~$\widetilde g_{\cent \qub_1}$ gives the effective transverse coupling strength and, thus, the system is not fully diagonalized unless~$\widetilde g_{\cent \qub_1}=0$. Moreover, one can show~\cite{Heunisch2023a} that the delocalization in the effective eigenstates also becomes minimal (but not completely suppressed) if~$\widetilde g_{\cent \qub_1}=0$. Thus, regarding the single-qubit gate operations the preferable idling configuration of the qubit, center mode, and coupler frequencies is such that it suppresses the effective coupling~$\widetilde g_{\cent \qub_1}$. In this paper, we assume that the tunable couplers have been designed in the two-island configuration, see Fig.~\ref{fig:fig1} and Ref.~\onlinecite{Marxer2023}. Consequently, the couplings $g_{\qub_1 \tc_1}$ and $g_{\cent \tc_1}$ have different signs, resulting in the idling frequency of the coupler lying below those of the qubit and the center element.

Due to the dressing, the qubit transition frequencies become dependent on the state of the center mode. In other words, the transverse XY interaction is transformed in the diagonalization process into a longitudinal ZZ coupling which is quantified in terms of the ZZ coupling strength defined as
\begin{equation}\label{eq:ZZ_coup}
    \zeta_{\cent\qub_1} = (\widetilde \omega_{11} - \widetilde \omega_{01}) - (\widetilde \omega_{10}-\widetilde \omega_{00}),
\end{equation}
where~$\widetilde \omega_{ n_{\qub_1}n_\cent}$ is the dressed eigenfrequency corresponding to the eigenstates~$\ket{n_{\qub_1}n_\cent }$, where~$n_{\qub_1},n_\cent = 0,1,2, \ldots$. We have chosen the labels here such that the overlap with the local state~$\ket{ n_{\qub_1}n_\cent}^0$ is maximal~\cite{Chu2021}. Here~$\ket{ n_{\qub_1}n_\cent}^0$ are the joint eigenstates of the local excitation number operators~$\hat n_{\qub_1}$ and~$\hat n_\cent$. 

During idling, any residual ZZ coupling is detrimental since it results in collection of conditional phase. It turns out that in general for the transmon-based coupler designs the delocalization and the residual collection of conditional phase, determined by~$\widetilde g_{\cent \qub_1}$ and~$\zeta_{\cent\qub_1}$ respectively, cannot be suppressed in the same configuration. Since the delocalization causes errors only during single-qubit gates, whereas the residual ZZ interaction leads to collection of conditional phase also during idling, one typically chooses to define the idling configuration by minimizing the ZZ coupling~\cite{Marxer2023}. Here, we follow this convention but also give a comprehensive analysis how our multi-mode coupler design also suppresses delocalization of the computational qubit states and thus results into improved single-qubit-gate accuracy compared to the more conventional square-grid implementations.

\begin{figure*}
    \centering
    \includegraphics[width=1.0\linewidth]{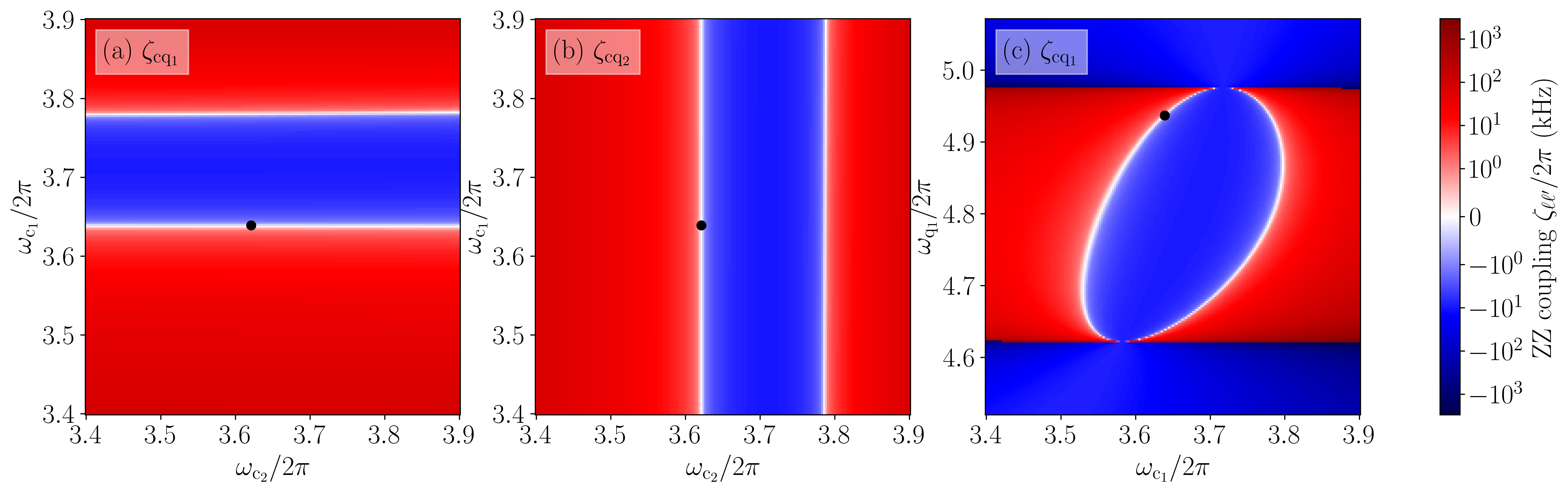}
    \caption{ZZ couplings~$\zeta_{\ell\ell'}$ in the minimal system as a function of the frequencies~$\omega_{\tc_1}$ and~$\omega_{\tc_2}$ of the tunable couplers~$\tc_1$ and~$\tc_2$. ZZ coupling between (a) the qubit~$\qub_1$ and the center mode~$\cent$,  (b) the qubit~$\qub_2$ and the center mode~$\cent$. In the chosen parameter range, the ZZ coupling between qubit~$\qub_j$ and the center mode has a negligible dependence on the tunable-coupler frequencies of the other qubits.
    (c) The ZZ coupling between the qubit~$\qub_1$ and the center mode~$\cent$ as a function of the frequencies~$\omega_{\tc_1}$ and~$\omega_{\qub_1}$. Black dots correspond to the chosen idling frequencies, see Tab.~\ref{tab:parameters}. The direct ZZ coupling between the qubits~$\qub_1$ and~$\qub_2$ is less than~\SI{10}{\kilo\hertz} for the chosen parameters. These results generalize for additional qubits.}
    \label{fig:zz_landscape}
\end{figure*}

Full calibration in the setup consisting of several qubits coupled to the center mode through tunable couplers becomes an exhaustive ordeal. In our numerical simulations, we apply an approximate but accurate search for the idling configuration that can be straightforwardly applied also in an experimental context. Instead of searching for the idling configuration of the whole system at once, we minimize the residual ZZ couplings for each center-mode--qubit pair separately, following the procedure presented above. We define the idling configuration such that~$\zeta_{\cent \qub_j} =0$ for each qubit with~$j=1,\ldots, N$. We show in Fig.~\ref{fig:zz_landscape} the calibration for the system with~$N=2$ qubits. 
In general, we choose to describe the system in the eigenbasis of the Hamiltonian defined in the above idling configuration. We define the labeling of the eigenstates as $\ket{n_{\qub_1}\ldots n_{\qub_N}n_\cent}$ such that the overlap with the local state $\ket{n_{\qub_1}\ldots n_{\qub_N}n_\cent}^0$ is maximal. Again, the states $\ket{n_{\qub_1}\ldots n_{\qub_N}n_{\cent}}^0$ are the joint eigenstates of the local excitation-number operators~$\hat n_{\qub_j}$ and~$\hat n_\cent$, where $j=1,\ldots,N$. The computational states relevant for universal quantum computation are then such eigenstates $\ket{n_{\qub_1}\ldots n_{\qub_N}n_\cent}$ that have $n_{\qub_i}=0$ or $n_{\qub_i}=1$, and $n_{\rm c} = 0$.

\begin{table}[t]
    \centering
    \begin{tabular}{c|c|c|c|c}
        Element & $\omega_\ell/2\pi$ (GHz) & $\alpha_\ell/2\pi$ (MHz) & $\beta_{\cent\ell}$ & $\beta_{\qub_j\tc_j}$\\
        \hline
        $\qub_1$ & 4.937 & -183 & 0.000842 & \\
        $\qub_2$ & 4.919 & -181 & 0.000882 & \\
        $\qub_3$ & 4.952 & -174 & 0.000861 & \\
        $\qub_4$ & 4.970 & -180 & 0.000904 & \\
        $\qub_5$ & 4.888 & -176 & 0.000862 & \\
        $\qub_6$ & 4.965 & -176 & 0.000838 & \\
        $\cent$ & 4.796 & -179 &  &  \\
        $\tc_1$ & 3.639 & -228 & -0.0145 & 0.0194\\
        $\tc_2$ & 3.621 & -228 & -0.0153 & 0.0193\\
        $\tc_3$ & 3.671 & -228 & -0.0149 & 0.0194 \\
        $\tc_4$ & 3.704 & -228 & -0.0157 & 0.0193\\
        $\tc_5$ & 3.602 & -228 & -0.0149 & 0.0194\\
        $\tc_6$ & 3.703 & -228 & -0.0144 & 0.0194
    \end{tabular}
    \caption{Basic parameters of the Hamiltonian in Eq.~\eqref{eq:ham1} used in this work. Negative signs in the couplings between the tunable couplers and the center mode set the idling frequencies of the couplers to frequencies below the qubits and the center mode~\cite{Marxer2023}. All the ZZ couplings in the system are less than~\SI{20}{\kilo\hertz}.}
    \label{tab:parameters}
\end{table}

\section{CZ gate realization}\label{sec:gate}
\begin{figure*}
    \centering
    \includegraphics[width=0.9\linewidth]{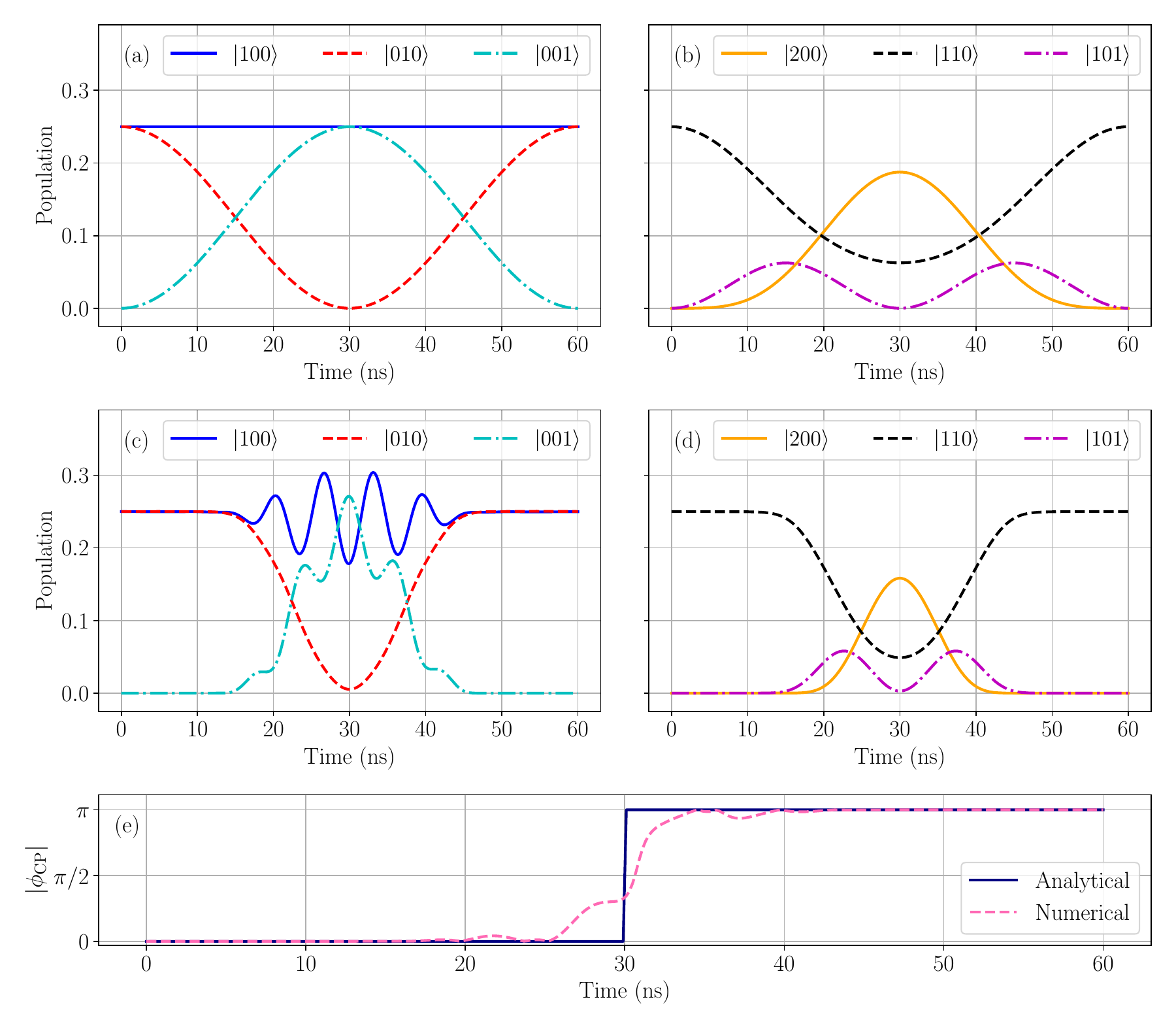}
    \caption{
    Population dynamics for the states performing the gate obtained with the simple model in Eq.~\eqref{eq:hamiltonians}, and with the complete Hamiltonian in Eq.~\eqref{eq:minimal_hamiltonian} including pulse shapes and tunable couplers. We use~$\ket{\psi(0)} = (\ket{000} + \ket{100} + \ket{010} + \ket{110})/2$ as the initial state, and the labeling convention is~$\ket{n_{\qub_1}n_{\qub_2}n_{\cent}}$. %Time evolution of the states performing the gate is presented. 
    (a) Single-excitation states of the analytical model. (b) Two-excitation states of the analytical model. (c) Single-excitation states of the complete system. (d) Two excitation states of the complete system. The CZ-gate infidelity in the full model is~$1-\mathcal{F} = 6.2\cdot10^{-7}$. In the complete system also the coupler states momentarily gain population (data not shown). (e) Accumulation of the conditional phase~$\phi_{\rm CP}$ during the gate using the analytical result in Eq.~\eqref{eq:analyt_cp} (solid dark blue) and the full numerical solution (dashed pink), which gives for the conditional phase~$\phi_{\rm CP}(\tau)=3.141$ at the end of the gate.}
    \label{fig:populations}
\end{figure*}    

\begin{figure}
    \centering
    \includegraphics[width=1.0\linewidth]{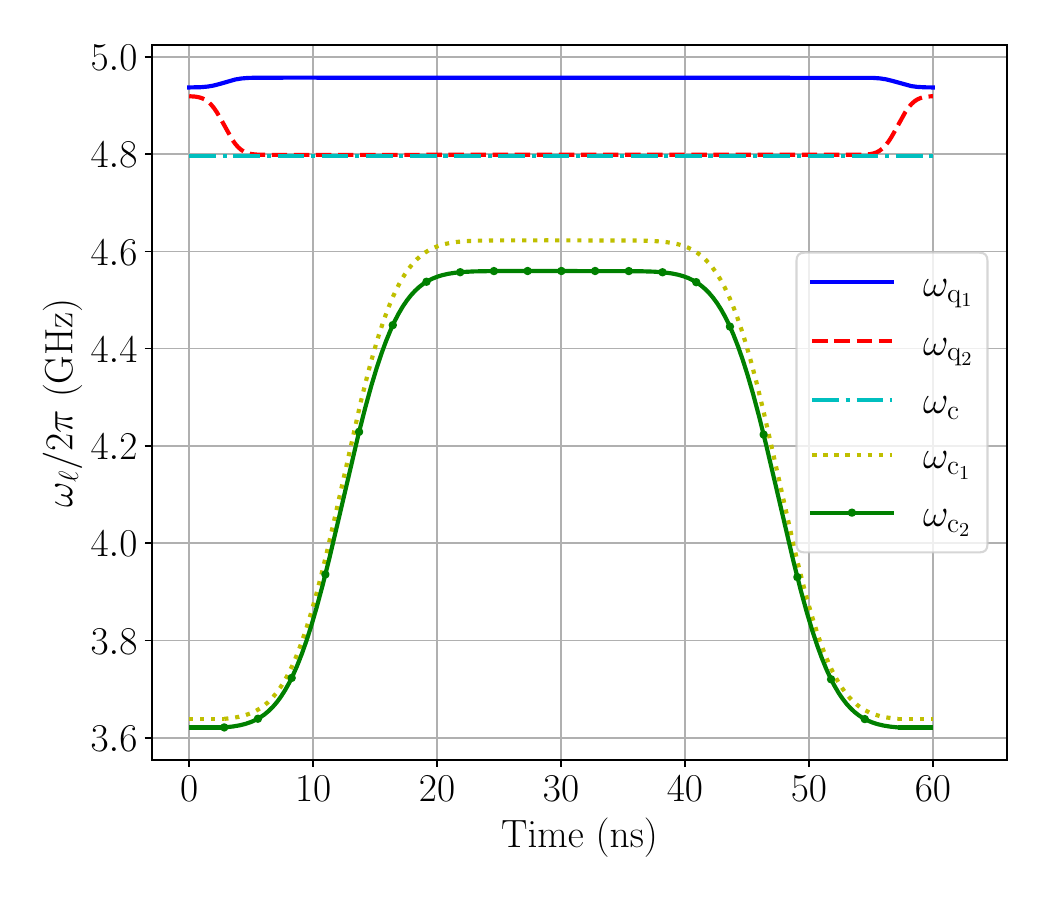}
    \caption{Schematic of the pulse schedule for a~$\tau=\SI{60.0}{\nano\second}$ pulse with~$\sigma_{\qub}=\SI{1.0}{\nano\second}$ and~$\sigma_{\tc}=\SI{3.4}{\nano\second}$. First the qubits~$\qub_1$ and~$\qub_2$ are pulsed to their respective operation points,~$\qub_2$ close to resonance with the center mode~$\cent$ and~$\qub_1$ anharmonicity~$|\alpha_{\qub_1}|$ above. Then the tunable couplers~$\tc_1$ and~$\tc_2$ are pulsed closer to the center mode~$\cent$. As predicted by Eq.~\eqref{eq:coupling_restriction}, the coupler~$\tc_1$ goes slightly higher than~$\tc_2$, generating a stronger effective coupling.}
    \label{fig:pulse}
\end{figure}

The discussion above draws the guidelines on how to determine the computational states such that the idling and single-qubit gate operations can be conducted with high accuracy. We assume that the quantum state of the multi-qubit system can be read out in the computational basis using the conventional dispersive readout techniques for transmons. Here, we concentrate on the implementation of a two-qubit CZ gate in our setup, and consider single-qubit gates in a following section. 

In a previous work, CZ gate between two peripheral qubits in the hexagonal unit cell has been realized with the so-called MOVE-CZ-MOVE protocol~\cite{Renger2025}. MOVE-CZ-MOVE consists of three successive operations for the system consisting of two peripheral qubits $\qub_1$ and $\qub_2$, and the center element~$\cent$. The first MOVE operation transfers the state of $\qub_2$ into the center element. This is followed by a CZ operation between the center element and $\qub_1$, realized using the diabatic gate scheme in which the conditional phase is generated by a full Rabi cycle between the two-excitation states $\ket{101}$ and $\ket{200}$~\cite{Li2020, Sung2021}. Finally, the state of the center element is transferred back to the qubit $\qub_2$ using another MOVE. Effectively, this creates a CZ gate between the peripheral qubits $\qub_1$ and $\qub_2$, ideally not influencing the spectating qubits in the QPU, and allowing for a tunable all-to-all connectivity within the unit cell of the honeycomb topology. The downside of this protocol is that the operations are done sequentially, resulting in a gate time given by $\tau = 2\tau_{\rm MOVE} + \tau_{\rm CZ}$, where $\tau_{\rm MOVE}$ and $\tau_{\rm CZ}$ are the durations of the MOVE and CZ operations, respectively.

Here, we reduce the gate duration by conducting the above operations simultaneously~\cite{Gu2021}. This means that we generate a pulse sequence for the peripheral qubits and their couplers such that the system experiences full Rabi oscillations in the single-excitation and two-excitation manifolds. This potentially decreases the duration of the effective CZ gate between the peripheral qubits.
In order to demonstrate our high-connectivity gate, let us first consider a system with two qubits~$\qub_1$ and~$\qub_2$, two tunable couplers~$\tc_1$ and~$\tc_2$, and the center mode $\cent$. The Hamiltonian of the system is given by
\begin{align}
    \hop/\hbar &=
    \sum_{\ell=\qub_1,\qub_2,\cent,\tc_1,\tc_2}\left[\omega_\ell \nop_\ell + \frac{\alpha_\ell}{2}\nop_\ell(\nop_\ell-1)\right]
    \nonumber\\
    &- \sum_{\ell=\qub_1,\qub_2,\tc_1,\tc_2}g_{\cent\ell}\left(\aop_\cent^\dag-\aop_\cent\right)\left(\aop_{\ell}^\dag-\aop_{\ell}\right)
    \nonumber\\
    &- \sum_{j=1}^2g_{\qub_j\tc_j}\left(\aop_{\qub_j}^\dag-\aop_{\qub_j}\right)\left(\aop_{\tc_j}^\dag-\aop_{\tc_j}\right).\label{eq:minimal_hamiltonian}
\end{align}
We describe our gate scheme first in terms of an approximative analytic model in which the gate is realized with square pulses applied to the participating qubits and couplers. Later, we test the protocol against a full numerical simulation of the Hamiltonian in Eq.~\eqref{eq:minimal_hamiltonian} with experimentally more relevant pulse shapes.

\subsection{Analytical model}\label{sec:analyt_mod}

Here, we decouple the tunable couplers from the rest of the system by performing the Schrieffer--Wolff transformation (see App.~\ref{sec:SF} for details), and obtain an effective Hamiltonian of the form
\begin{align}
    \hop_{\rm SW}/\hbar &=
    \sum_{\ell=\qub_1,\qub_2,\cent,\tc_1,\tc_2}\left[\widetilde\omega_\ell \nop_\ell + \frac{\alpha_\ell}{2}\nop_\ell(\nop_\ell-1)\right]
    \nonumber\\
    &+\sum_{j=1}^2\widetilde g_{\cent\qub_j}\left(\aop_\cent^\dag\aop_{\qub_j} + \aop_\cent\aop_{\qub_j}^\dag\right)
    \nonumber\\
    &+\widetilde g_{\tc_1\tc_2}\left(\aop_{\tc_1}^\dag\aop_{\tc_2} + \aop_{\tc_1}\aop_{\tc_2}^\dag\right),\label{eq:SF_ham}
\end{align}
where we have also performed the rotating wave approximation on the couplings. We observe that after the transformation the couplers are decoupled from the qubits and the center mode and, thus, we can discard the tunable coupler degrees of freedom and focus only on the subsystem of two qubits and the center element. The coupling strengths~$\widetilde g_{\cent\qub_j}$ here depend on the frequencies of the tunable couplers and, in particular, can be set to zero, which we for simplicity choose as the idling configuration for this approximative model. A gate can then be performed by tuning the coupler frequencies to desired operation values, which create effective couplings between the qubits and the center mode. Notice that in the numerical simulations, described in the following section, we choose the idling value for~$\widetilde g_{\cent\qub_j}$ such that it minimizes the ZZ-coupling, and use the eigenstates of Eq.~\eqref{eq:SF_ham} as the computational states, as discussed in Sec.~\ref{sec:idling}.

Without a loss of generality, we assume that~$\widetilde\omega_{\qub_1} >\widetilde\omega_{\qub_2},\widetilde\omega_{\cent}$, and label the computational states as~$\ket{n_{\qub_1}n_{\qub_2}n_{\cent}}$ as described in the previous section. The essential states for the process are~$\ket{001}$,~$\ket{010}$ and~$\ket{100}$ in the single-excitation manifold, and~$\ket{200}$,~$\ket{110}$ and~$\ket{101}$ in the two-excitation manifold. Projecting the Hamiltonian in Eq.~\eqref{eq:SF_ham} to these subspaces, we obtain the matrix representation~$\hop_{\rm SW} \rightarrow \hop_1 \oplus \hop_2 $ where
\begin{align}
    \hop_1 &\equalhat 
    \hbar
    \begin{pmatrix}
        \widetilde\omega_\cent & \widetilde g_{\cent\qub_2} & \widetilde g_{\cent\qub_1}\\
        \widetilde g_{\cent\qub_2} & \widetilde\omega_{\qub_2} & 0\\
        \widetilde g_{\cent\qub_1} & 0 & \widetilde\omega_{\qub_1}
    \end{pmatrix},
    \nonumber\\
    \hop_2 &\equalhat
    \hbar
    \begin{pmatrix}
        2\widetilde\omega_{\qub_1} +\alpha_{\qub_1} & 0 & \sqrt{2}\widetilde g_{\cent\qub_1}\\
        0 & \widetilde\omega_{\qub_1} + \widetilde\omega_{\qub_2} & \widetilde g_{\cent \qub_2}\\
        \sqrt{2}\widetilde g_{\cent\qub_1} & \widetilde g_{\cent \qub_2} & \widetilde\omega_{\qub_1} + \widetilde\omega_{\cent}
    \end{pmatrix}.\label{eq:hamiltonians}
\end{align}
We apply square pulses to the qubits such that during the gate~$\widetilde\omega_{\qub_2} = \widetilde\omega_{\cent}$ and~$\widetilde\omega_{\qub_1} + \alpha_{\qub_1} = \widetilde\omega_{\qub_2}$, so that all the three states in the two excitation manifold are in resonance. In the single-excitation manifold the states~$\ket{010}$ and~$\ket{001}$ are also in resonance, but the state~$\ket{100}$ is not. In order to proceed analytically we assume that~$\widetilde g_{\cent\qub_1}\ll \widetilde\omega_{\qub_1} - \widetilde\omega_\cent $ and, consequently, neglect the effect of $g_{\cent\qub_1}$ in the single-excitation manifold. More accurate but less intuitive analysis using Jacobian diagonalization~\cite{Jacobi1846, Tuorila2017} is given in App.~\ref{sec:Jacobi}.

Since we have made the rotating-wave approximation, the dynamics in single- and two-excitation manifolds is decoupled from the other manifolds, and is described by the propagator $\hat U(t) = \hat U_1(t)\oplus \hat U_2(t)$ where
\begin{widetext}
    \begin{align}
        \uop_1(t) &\equalhat
        \begin{pmatrix}
            e^{-\ii \widetilde\omega_{q_2}t}\cos(\widetilde g_{\cent \qub_2}t) & -\ii e^{-\ii \widetilde\omega_{q_2}t}\sin(\widetilde g_{\cent \qub_2}t) & 0\\
            -\ii e^{-\ii \widetilde\omega_{q_2}t}\sin(\widetilde g_{\cent \qub_2}t) & e^{-\ii \widetilde\omega_{q_2}t}\cos(\widetilde g_{\cent \qub_2}t) & 0\\
            0 & 0 & e^{-\ii \widetilde\omega_{q_1}t}\\
        \end{pmatrix},
        \nonumber
        \\
        \uop_2(t) &\equalhat
        \frac{e^{-\ii (\widetilde\omega_{\qub_1}+\widetilde\omega_{\qub_2}) t}}{\Omega^2}
        \begin{pmatrix}
            \widetilde g_{\cent\qub_2}^2+2\widetilde g_{\cent\qub_1}^2\cos(\Omega t)
            &
            \sqrt{2}\widetilde g_{\cent\qub_1} \widetilde g_{\cent\qub_2}[\cos(\Omega t) - 1]
            &
            -\ii \Omega\sqrt{2}\widetilde g_{\cent\qub_1}
            \sin(\Omega t)
            \\
             \sqrt{2}\widetilde g_{\cent\qub_1} \widetilde g_{\cent\qub_2}[\cos(\Omega t) - 1]
            &
            2\widetilde g_{\cent\qub_1}^2+\widetilde g_{\cent\qub_2}^2\cos(\Omega t)
            &
            -\ii \Omega \widetilde g_{\cent\qub_2}
            \sin(\Omega t)
            \\
            -\ii \Omega\sqrt{2}\widetilde g_{\cent\qub_1}
            \sin(\Omega t)
            &
             -\ii \Omega \widetilde g_{\cent\qub_2}
            \sin(\Omega t)
            &
            \Omega^2\cos(\Omega t)
        \end{pmatrix},\label{eq:propagators}
    \end{align}
\end{widetext}
and we have defined the oscillation rate in the two excitation manifold as
\begin{equation}\label{eq:two_exc_omega}
    \Omega\equiv \sqrt{2\widetilde g_{\cent\qub_1}^2 + \widetilde g_{\cent\qub_2}^2}.
\end{equation}
In this two-qubit system the computational states are~$\ket{000}$,~$\ket{100}$,~$\ket{010}$ and~$\ket{110}$, so we observe that there is population transfer out of this computational subspace in both single- and two-excitation manifolds. For a good fidelity quantum gate it is required that the population returns to the computational subspace at the end of the gate.
In the two-excitation manifold this occurs only if~$\Omega t = 2\pi n$, and in the single-excitation manifold if~$\widetilde g_{\cent\qub_2}t = n\pi$, with an integer~$n$. We choose~$n=1$ to minimize the gate duration, and using Eq.~\eqref{eq:two_exc_omega}, we observe that full population recovery in both manifolds is obtained if
\begin{equation}\label{eq:coupling_restriction}
    \widetilde g_{\cent\qub_1} = \pm\sqrt{\frac{3}{2}}\widetilde g_{\cent\qub_2},\quad \Omega = 2|\widetilde g_{\cent\qub_2}|.
\end{equation}
The gate is demonstrated in Fig.~\ref{fig:populations}~(a) and~(b). The system is initially in an equal superposition of the computational states, and performs a full Rabi oscillation between the single-excitation states~$\ket{010}$ and~$\ket{001}$. Simultaneously the three two-excitation states perform their own oscillations. The oscillation rate of the state involving the center-mode depends on the state of the higher-frequency qubit, such that the state~$\ket{101}$ oscillates twice as fast as the state~$\ket{001}$. The~$\ket{200}$ never becomes fully populated, since the couplings~$\sqrt{2}\widetilde g_{\cent\qub_1}$ and~$\widetilde g_{\cent\qub_2}$ are not equal. 

For a working CZ gate we also need to make sure that the gate can give a~$\pi$ conditional phase. Setting the system initially in a general state
\begin{equation}
    \ket{\psi(0)} = \left(c_{00}\ket{00} + c_{10}\ket{10} + c_{01}\ket{01} + c_{11}\ket{11}\right)\otimes\ket{0},
\end{equation}
where the coefficients~$c_{ij}$ are complex numbers subject to normalization condition, and the center mode is in the ground state. Time evolution is determined by the propagators in Eq.~\eqref{eq:propagators} with the gate duration set to~$\tau = 2\pi/\Omega$, $\ket{\psi(\tau)}=\uop(\tau)\ket{\psi(0)}$. Tracing out the center element degree of freedom then gives a two-qubit density matrix with the elements (indexing the states as~$\ket{n_{\qub_1}n_{\qub_2}}$)
\begin{align}
    \bra{00}\dens_{\rm 2q}(\tau)\ket{01} &= 
    |c_{00}||c_{01}|
    e^{\ii (\widetilde\omega_{\qub_2}\tau + \vartheta_{00} - \vartheta_{01})}
    \cos(\widetilde g_{\cent\qub_2}\tau),
    \nonumber
    \\
    \bra{00}\dens_{\rm 2q}(\tau)\ket{10} &= 
    |c_{00}||c_{10}|
    e^{\ii (\widetilde\omega_{\qub_1}\tau + \vartheta_{00} - \vartheta_{10})},
    \nonumber
    \\
    \bra{00}\dens_{\rm 2q}(\tau)\ket{11} &=
    |c_{00}||c_{11}|
    e^{\ii [(\widetilde\omega_{\qub_1}+\widetilde\omega_{\qub_2})\tau + \vartheta_{00} - \vartheta_{11}]}.
\end{align}
where we have written the complex numbers~$c_{jk}$ in the polar form,~$c_{jk} \equiv |c_{jk}|e^{\ii \vartheta_{jk}}$. The single- and two-qubit phases collected during the gate are obtained from these matrix elements as~$\phi_{jk}=\arg\left[\bra{00}\dens_{\rm 2q}\ket{jk}\right]$,
\begin{align}
    \phi_{01}(\tau) &=
    \widetilde\omega_{\qub_2}\tau+\vartheta_{00}-\vartheta_{01}
    -\frac{\pi}{2}\Big\{1-\sgn\Big[\cos(\widetilde g_{\cent\qub_2}\tau)\Big]\Big\},
    \nonumber
    \\
    \phi_{10}(\tau) &=
    \widetilde\omega_{\qub_1}\tau+\vartheta_{00}-\vartheta_{10},
    \nonumber
    \\
    \phi_{11}(\tau) &=
    (\widetilde\omega_{\qub_1}+\widetilde\omega_{\qub_2})\tau+\vartheta_{00}-\vartheta_{11},
\end{align}
where~$\sgn$ is the sign function. The conditional phase is defined, similar to the ZZ-coupling in Eq.~\eqref{eq:ZZ_coup}, as the difference between the two qubit phase~$\phi_{11}$ and the single-qubit phases~$\phi_{10}$ and~$\phi_{01}$ as
\begin{align}
    \phi_{\rm CP}(\tau) \equiv&
    \phi_{11}(\tau) -\phi_{10}(\tau)-\phi_{01}(\tau)
    \nonumber\\
    =&
    \frac{\pi}{2}\Big\{1-\sgn\Big[\cos(\widetilde g_{\cent\qub_2}\tau)\Big]\Big\}\nonumber\\
    &-\Big(
    \vartheta_{11}
    -\vartheta_{10}
    -\vartheta_{01}
    +\vartheta_{00}
    \Big).\label{eq:analyt_cp}
\end{align}
Thus, we observe that if~$\widetilde g_{\cent\qub_2}\tau = \pi$, we obtain the desired change of the conditional phase by~$\pi$. However, unlike in the conventional CZ gate where the conditional phase originates from the Rabi oscillations in the two excitation manifold~\cite{Li2020, Sung2021}, here it actually arises from the oscillation in the single-excitation manifold, between the lower frequency qubit and the center element.

In a frame rotating with the qubit frequencies~$\widetilde\omega_{\qub_{1,2}}$ we thus write the gate unitary acting on the two qubit system (state ordering~$\{\ket{00},\,\ket{01},\,\ket{10},\,\ket{11}\}$) as
\begin{equation}\label{eq:almost_cz}
    \uop_{\qub_1\qub_2} \equalhat
    \begin{pmatrix}
        1 & 0 & 0 & 0\\
        0 & -1 & 0 & 0\\
        0 & 0 & 1 & 0\\
        0 & 0 & 0 & 1
    \end{pmatrix},
\end{equation}
which is not exactly a CZ gate. However, the CZ gate is decomposed as a product of a~$\pi$-phase gate of qubit~$\qub_2$ and the above gate as
\begin{equation}
    \uop_{\rm CZ} 
    \equalhat
    \begin{pmatrix}
        1 & 0 & 0 & 0\\
        0 & -1 & 0 & 0\\
        0 & 0 & 1 & 0\\
        0 & 0 & 0 & -1
    \end{pmatrix}
    \begin{pmatrix}
        1 & 0 & 0 & 0\\
        0 & -1 & 0 & 0\\
        0 & 0 & 1 & 0\\
        0 & 0 & 0 & 1
    \end{pmatrix}.
\end{equation}
Our gate scheme thus first creates a conditional phase into single-excitation state of the qubit~$\qub_2$, and then transfers it into the conditional phase of the doubly-excited state by changing the phase of the qubit~$\qub_2$ by~$\pi$. In App.~\ref{app:mczm} we perform a similar analysis for the MOVE-CZ-MOVE gate and show that the effective gate unitary is the same as in Eq.~\eqref{eq:almost_cz}. However, the actual unitary propagators differ from those in Eq.~\eqref{eq:propagators}, meaning that the physical origin in these two gate schemes is distinct.

There are two couplings involved in the gate, of which~$\widetilde g_{\cent\qub_1}$ is stronger, due to the relation in Eq.~\eqref{eq:coupling_restriction}. However, in the MOVE-CZ-MOVE there is no such restriction, and we could assume that both MOVE and CZ gates can be performed with the larger coupling strength. The total gate duration of the MOVE-CZ-MOVE protocol is
\begin{equation}
    \tau_{\rm 2MOVE + CZ} = \frac{\sqrt{2}+1}{\sqrt{2}}\frac{\pi}{\widetilde g_{\cent\qub_1}}.
\end{equation}
Our new implementation instead has the duration~$\tau = \sqrt{3/2}\pi/\widetilde g_{\cent\qub_1}$, so that
\begin{equation}
    \frac{\tau_{\rm 2MOVE+CZ}}{\tau} = \frac{\sqrt{2}+1}{\sqrt{3}}\approx\sqrt{2},
\end{equation}
meaning that our gate is roughly a factor of~$\sqrt{2}$ faster than the MOVE-CZ-MOVE protocol.

\subsection{Numerical optimization of the CZ gate}\label{sec:gate_optim}

In the above analysis, we made several approximations which may not hold in realistic systems. These approximations include the decoupling of the tunable couplers, ignoring the off-resonant Rabi oscillations with the higher frequency qubit in the single-excitation manifold, as well as discarding the counter-rotating terms in the couplings. We also assumed square-pulse shapes for the qubits and the couplers. Here, we numerically demonstrate the realization of the above-described gate protocol using the full Hamiltonian in Eq.~\eqref{eq:minimal_hamiltonian}, without making any of the above mentioned approximations. We set the system into its idling configuration where qubit~$\qub_1$ and~$\qub_2$ frequencies are detuned roughly by their respective anharmonicities above the frequency of the center element. The tunable couplers are below the center element in frequency and their respective idling frequencies are determined by minimizing the ZZ-coupling, as described in Sec.~\ref{sec:system}. 

In order to perform the gate we apply a time-dependent flux pulse on one of the qubits, say~$\qub_2$, which takes it down in frequency close to resonance with the center mode. Simultaneously, we also apply a flux pulse on qubit~$\qub_1$, which keeps it close to its original frequency so that the states~$\ket{110}$ and~$\ket{200}$ are close to resonance. The latter qubit flux pulse provides more freedom to calibrate the gate for higher fidelity, as well as greater flexibility in choosing the idling frequency configuration for the qubits~\cite{marxer2025}. After both qubits are swept to their respective operation points, flux pulses are applied on both tunable couplers, taking their frequencies closer to that of the center mode. This effectively produces the desired coupling strengths~$\widetilde g_{\cent\qub_1}$ and~$\widetilde g_{\cent\qub_2}$ in~Eq.~\eqref{eq:SF_ham}. The tunable coupler~$\tc_1$ must be taken slightly closer to the center mode than~$\tc_2$, due to the condition defined in Eq.~\eqref{eq:coupling_restriction}. The closer the couplers are to the center mode, the stronger are the effective coupling strengths, and thus the faster the gate. However, too strong couplings can induce coherent errors such as leakage out of the computational subspace. An example of the pulse schedule is shown in Fig.~\ref{fig:pulse}, where we show the time-dependent frequencies of the individual transmons during the gate.

We use flat-top Gaussian (Gaussian-filtered square pulse) shape for all flux pulses, defined as
\begin{equation}
    f(t, A, \sigma) = \frac{A}{2} \left[\erf\left(\frac{t-\tau_{\rm b}}{\sqrt{2} \sigma}\right)
    -\erf\left(\frac{t-\tau_{\rm c}-\tau_{\rm b}}{\sqrt{2} \sigma}\right) \right],
\end{equation}
where~$\erf$ is the error function,~$A$ and~$\tau_{\rm c}$ are the amplitude and duration of the square pulse,~$\sigma$ is the width of the Gaussian filter, and the buffer time~$\tau_{\rm b}=2\sqrt{2}\sigma$ sets the rise time of the pulse. The total duration of the pulse is then~$\tau = \tau_{\rm c} + 2\tau_{\rm b}$. The frequencies of the qubit and coupler modes in the Hamiltonian in Eq.~\eqref{eq:minimal_hamiltonian} are thus time dependent,
\begin{equation}
    \omega_{\ell}(t) = \omega_{\ell}^{\text{idle}} + f(t, A_{\ell}, \sigma_\ell),\quad\ell = \qub_{1,2},\tc_{1,2},
\end{equation}
so that the operation frequency is determined by the amplitude,~$\omega_{\ell}^{\text{oper}} = \omega_\ell^{\text{idle}} + A_\ell$. The coupling strengths between the local modes also become time dependent, due to their dependence on the frequency~\cite{Yan2018}, defined as
\begin{equation}
    g_{\ell\ell'}(t) = \beta_{\ell\ell'}\sqrt{\omega_\ell(t)\omega_{\ell'}(t)},
\end{equation}
where~$\beta_{\ell\ell'}$ is a dimensionless parameter determined by the capacitance matrix of the whole circuit. 

We have multiple free parameters which allow a good amount of tunability for the optimization of our gate scheme. For simplicity we set~$\sigma_{\qub}=\SI{1.0}{\nano\second}$ for both qubit pulses, and begin and end the coupler pulses one buffer time~$\tau_{\text{b}}=2\sqrt{2}\sigma_{\qub}$ after the start and before the end of the qubit pulses, respectively (see Fig.~\ref{fig:pulse}). The remaining free parameters are the four amplitudes~$A_{\qub_1}$,~$A_{\qub_2}$,~$A_{\tc_1}$ and~$A_{\tc_2}$, and the width~$\sigma_{\tc}$ of the Gaussian filter, which we choose to be equal for both coupler pulses. We could in principle also optimize the duration~$\tau$ of the pulse scheme, but we instead show here data obtained with a fixed duration of~$\tau=\SI{60.0}{\nano\second}$.

We here perform a numerical optimization to find the set of parameters~$\{A_{\qub_1},A_{\qub_2},A_{\tc_1},A_{\tc_2},\sigma_{\tc}\}$ that gives the best gate fidelity. We first set the system into an initial state~$\ket{\psi(0)}$ which is some linear combination of the eigenstates of the idling Hamiltonian. We then apply the pulses as described above, and obtain some final state~$\dens(\tau)$. Since we calculate the dynamics in the laboratory frame, we need to remove the single qubit phases from the state. These can be calculated directly from the state, and removed with suitable single-qubit phase gates. The ideal state is obtained by applying the CZ gate (expressed in the full system Hilbert space) on the initial state,
\begin{equation}
    \ket{\psi^{\rm ideal}} = \uop_{\rm CZ}\ket{\psi(0)}.
\end{equation}
The ideal gate produces a pure state, so gate fidelity between the numerically obtained state and the ideal state is
\begin{equation}\label{eq:2Q_Fid}
    \mathcal{F} = 
    \bra{\psi^{\rm ideal}}\dens(\tau)\ket{\psi^{\rm ideal}}.
\end{equation}
We use the infidelity~$1-\mathcal{F}$ as a cost function, and obtain the optimal pulse parameters for the high-fidelity gate using the minimization routines from SciPy~\cite{Scipy2020}.

The time evolution of the population of the states participating in the gate is shown in Fig.~\ref{fig:populations} (c) and (d) for the initial state $\ket{\psi(0)} = (\hat I + \aop_{\qub_1}^\dag)(\hat I + \aop_{\qub_2}^\dag)\ket{\text{vac}}/2$, where~$\ket{\text{vac}}$ is the vacuum state of the system. We optimize the four amplitudes and the coupler-pulse rise time for a fixed duration of~$\tau=\SI{60.0}{\nano\second}$, and obtain the gate fidelity of~$\mathcal{F}\approx 6.2\cdot10^{-7}$. Comparing the populations to those obtained with the simple model in Fig.~\ref{fig:populations}~(a) and~(b) shows that both models are qualitatively similar. The largest difference is in the off resonant oscillations in the single-excitation manifold which were ignored in the simple model. The difference in shapes arises because in the simple model the couplings are created with square pulses whereas in the full system they are turned on gradually using the pulses displayed in Fig.~\ref{fig:pulse}.

In Fig.~\ref{fig:populations} we also show the accumulation of the conditional phase~$\phi_{\rm CP}$ during the gate. The analytical result in Eq.~\eqref{eq:analyt_cp} predicts a sharp step function change, caused by the change of the sign of cosine function in the middle of the gate. The full numerical simulations shows a more gradual change towards the desired value of~$\phi_{\rm CP}=\pi$.

\section{Spectator and hybridization induced errors}\label{sec:spectator_hybrid}
\begin{figure*}
    \centering
    \includegraphics[width=1.0\linewidth]{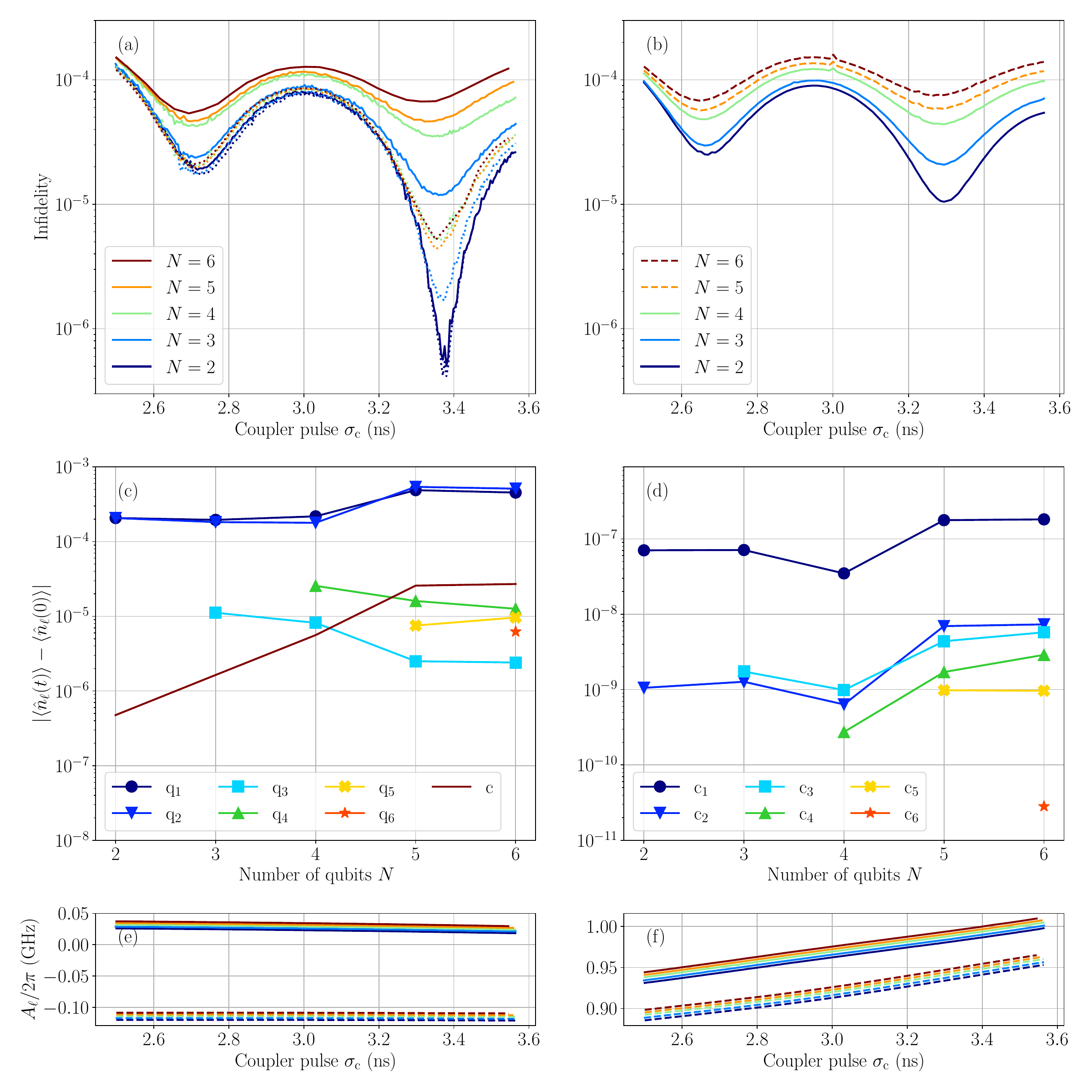}
    \caption{Effect of the number of qubits~$N$ in the system on the CZ gate fidelity as a function of coupler-pulse filter width~$\sigma_{\tc}$. We have used~$\sigma_{\qub}=\SI{1.0}{\nano\second}$ and the gate duration~$\tau=\SI{60.0}{\nano\second}$. (a) Gate infidelity (solid lines) and leakage error (dotted lines) for the initial state $\ket{\psi(0)} = (\hat I + \aop_{\qub_1}^\dag)(\hat I + \aop_{\qub_2}^\dag)\ket{\text{vac}}/2$. (b) Average gate infidelity.
    Dashed lines show a linear extrapolation of the existing data as a function of the number of qubits, for each value of~$\sigma_\tc$, since the numerical optimization of the average gate fidelity becomes computationally expensive for larger systems.
    (c) Qubit occupation difference at the beginning and at the end of the gate as a function of the number $N$ of qubits. For each $N$, we have used~$\sigma_{\tc}$ giving the smallest infidelity. (d) Coupler occupation difference at the beginning and the end of the gate as a function of the qubit number. (e)~Optimized amplitudes for~$\qub_1$ (solid) and~$\qub_2$ (dashed) for different system sizes, colours defined in the labels of~(a). (f)~Optimized amplitudes for~$\tc_1$ (solid) and~$\tc_2$ (dashed) for different system sizes.}
    \label{fig:spectators}
\end{figure*}

\begin{figure}
    \centering
    \includegraphics[width=1.0\linewidth]{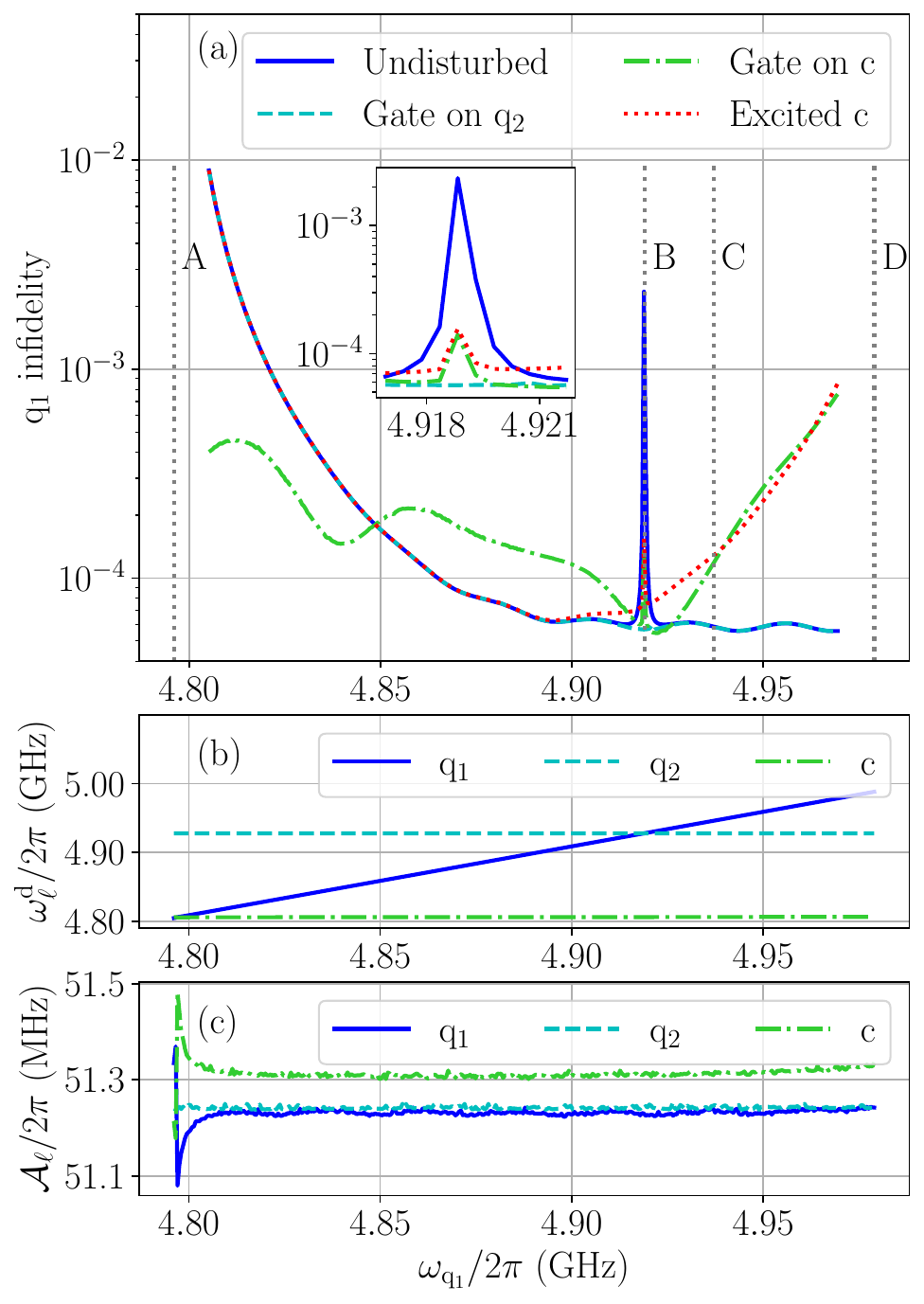}
    \caption{(a)~Single-qubit~$\pi$ gate infidelity of~$\qub_1$ as a function of~$\omega_{\qub_1}$. The system is initially in the vacuum state~$\ket{\psi(0)}=\ket{\text{vac}}$. In the undisturbed case single-qubit gate is applied only on~$\qub_1$ (solid blue). Dashed light blue line shows the situation when another gate is simultaneously applied on another qubit~$\qub_2$. Similar situation with additional gate applied to the center mode~$\cent$ is shown with the dash-dotted green curve. In the last case, we apply the gate only on~$\qub_1$, but the initial state is such that the center mode is excited, i.e. ~$\ket{\psi(0)}=\aop_\cent^\dag\ket{\text{vac}}$ (dotted red). Grey dashed lines~A,~B and~D denote the frequencies~$\omega_{\cent}/2\pi$,~$\omega_{\qub_2}/2\pi$ and~$(\omega_{\cent} + |\alpha_{\qub_1}|)/2\pi$, respectively. The line~C is the frequency of~$\qub_1$ taken from Tab.~\ref{tab:parameters}. The inset shows a zoom-in around the peak close to the line~B. The peak is not exactly at~$\omega_{\qub_2}$, since the hybridization of the computational states shifts the resonance frequencies slightly. (b)~Optimized drive frequency of each element. (c)~Optimized drive amplitude of each element.}
    \label{fig:sqg}
\end{figure}

\begin{figure*}
    \centering
    \includegraphics[width=0.85\linewidth]{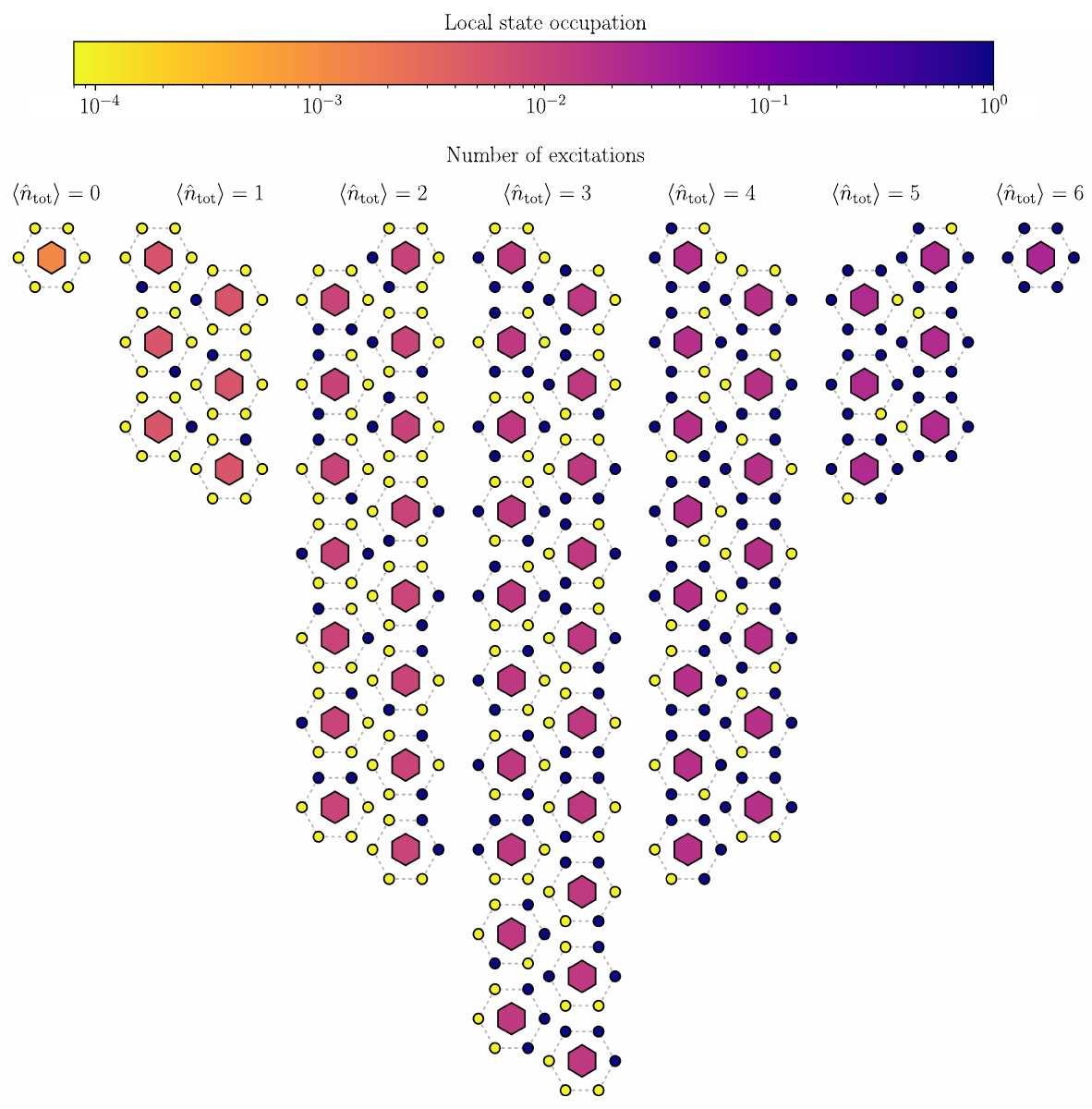}
    \caption{Occupations of the local modes in all the computational states, grouped according to the total number of excitations in the state,~$\nop_{\rm tot} = \sum_{\ell}\nop_\ell$. The strongest hybridization is between the qubit modes and the couplers, demonstrating that the coupling scheme prevents the qubits from hybridizing with each other.}
    \label{fig:compstates}
\end{figure*}

In the previous section, we studied a simple system with only two qubits. In realistic devices, however, the operating qubits are connected to other qubits which are not participating in the gate. Such spectator qubits introduce additional error channels which can decrease the fidelity of the gate. In this section we study how they affect our gate scheme.

\subsection{Two-qubit gates}
The unit cell of the honeycomb lattice consists of six qubits connected to a center mode with six tunable couplers. The CZ gate can be performed between any pair of these qubits. During the gate the remaining four qubits are idling. System parameters used in our numerical simulations are shown in Tab.~\ref{tab:parameters}. We perform the CZ gate between the qubits~$\qub_1$ and~$\qub_2$ such that~$\qub_2$ is pulsed close to resonance with the center mode~$\cent$, and optimize the four pulse amplitudes as a function of the coupler-pulse filter width~$\sigma_{\tc}$. In Fig.~\ref{fig:spectators}, we show the simulation data for the gate infidelity for a single initial state, as well as the average gate infidelity (see App.~\ref{sec:AVGF}). We then individually add one spectator qubit at a time, redo the optimization, and compute the gate infidelities for every added spectator. We observe that the error increases more or less steadily as the number of spectators grows, since each of them introduces an additional leakage channel. 
We also show the total population in the non-computational states with dashed lines in Fig.~\ref{fig:spectators}~(a). In the absence of spectators, this leakage error matches the gate error very closely. As additional spectators are added, the leakage error increases, but remains smaller than the total error, indicating that most of the gate error arises from population swapping between the qubits rather than leakage outside the computational subspace.

For Fig.~\ref{fig:spectators}~(c) and~(d) we choose the coupler-pulse filter width~$\sigma_{\tc}$ which gives the minimum infidelity, and show the population difference between the initial and final states of each element. We observe that the primary limiting factor is the population not returning to the operating qubits, but instead their populations differ from the initial values roughly in the range~$2\cdot10^{-4}-6\cdot10^{-4}$, implying that they swap population with each other. The effect of population staying in the center mode and leakage to spectators is considerably lower, but still enough to cause the observed increase in the error. Leakage to the tunable couplers is several orders of magnitude smaller. The leakage to the spectators is explained by the fact that when the operating couplers~$\tc_1$ and~$\tc_2$ are tuned to their operation frequencies, small unwanted transverse couplings are also introduced between the center mode and the spectators, since the computational states are the eigenstates of the idling Hamiltonian. This results into off-resonant population swapping. It could be possible to mitigate these spectator errors by simultaneously adjusting also the coupler frequencies of the spectators, so that the transverse coupling would be zero~\cite{zajac2021}. However, this can lead to non-zero ZZ coupling between the spectator and the center element, which potentially causes other problems. 

The optimized pulse amplitudes for the qubits and the couplers are shown in Fig.~\ref{fig:spectators}~(e) and~(f), respectively. 

\subsection{Single-qubit gates}
We also study how single-qubit gates perform in the system with the center mode. We implement the single-qubit gate on the element~$\ell$ as a microwave pulse, described by the Hamiltonian
\begin{align}\label{eq:sqg_hamiltonian}
    \hop_{\ell}^{\text{d}}(t) = \ii \hbar \mathcal{A}_\ell(t)\cos\left(\omega_{\ell}^{\text{d}}t + \phi_\ell\right)\left(\aop_\ell^\dag - \aop_\ell\right),
\end{align}
where~$\omega_{\ell}^{\text{d}}$ is the driving (angular) frequency and~$\phi_{\ell}$ is a phase that can be freely chosen to adjust the rotation axis of the single-qubit gate. To keep things simple we choose a Gaussian pulse envelope,
\begin{equation}
    \mathcal{A}_\ell(t) = \mathcal{A}_\ell 
    \exp\left[-\frac{(t-\tau/2)^2}{2\varsigma_\ell^2}\right]
    %e^{-\frac{(t-\tau/2)^2}{2\varsigma_\ell^2}},
\end{equation}
where~$\mathcal{A}_\ell$ is the amplitude,~$\tau$ is the duration of the pulse, and~$\varsigma_\ell$ determines the width of the envelope. For a chosen duration, the amplitude~$\mathcal{A}_\ell$ determines the rotation angle~$\theta_\ell$, which for weak driving is approximately
\begin{equation}\label{eq:theta}
    \theta_\ell \approx \int_0^\tau\D t \mathcal{A}_\ell(t)
    =
    \varsigma_\ell\mathcal{A}_\ell\sqrt{2\pi}\erf\left(\frac{\tau}{2\sqrt{2}\varsigma_\ell}\right).
\end{equation}
We choose our single-qubit gate length~$\tau=\SI{20.0}{\nano\second}$, and set~$\varsigma_\ell=\SI{4.0}{\nano\second}$. We study single-qubit gates with~$\theta_\ell = \pi$ and~$\phi_\ell = 0$. 

The ideal gate which the Hamiltonian in Eq.~\eqref{eq:sqg_hamiltonian} attempts to emulate is
\begin{equation}
    \uop_{\theta,\phi} \equalhat
    \begin{pmatrix}
        \cos\left(\frac{\theta}{2}\right) & -e^{\ii\phi}\sin\left(\frac{\theta}{2}\right) \\
        e^{-\ii\phi}\sin\left(\frac{\theta}{2}\right) & \cos\left(\frac{\theta}{2}\right)
    \end{pmatrix}.
\end{equation}
In order to perform the gate with high fidelity, we optimize the amplitude and the driving frequency. As an initial guess, we use~$\omega_{\ell}^{\text{drive}}=\omega_\ell$ and~$\mathcal{A}_\ell$ given by Eq.~\eqref{eq:theta}, and perform 2D Golden-section search~\cite{Rani2019} to find the optimal values. We study again the minimal system consisting of two qubits~$\qub_1$ and~$\qub_2$, their respective tunable couplers and the center mode~$\cent$.

Note that the microwave drive in Eq.~\eqref{eq:sqg_hamiltonian} couples to the local-transmon charge operator, whereas the computational modes are some linear combinations of the local-transmon eigenstates. This opens a new source of error, hybridization crosstalk, often overlooked in single-qubit gate calibration due to its small contribution to the total amount of crosstalk. However, as the errors caused by classical crosstalk can be reduced with novel pulse-shaping methods and developing the calibration, hybridization crosstalk can potentially become the dominating source of crosstalk error.

In Fig.~\ref{fig:sqg}~(a) we show how the infidelity of the single-qubit gate of qubit~$\qub_1$ depends on its frequency and other parallel operations nearby. Since we are only interested in the gate fidelity on qubit~$\qub_1$, we calculate the fidelity of a reduced state obtained by tracing out all the other degrees of freedom. Otherwise, the total fidelity would also include contributions from the parallel operations, making the comparison to a situation with a gate only on qubit~$\qub_1$ inconsistent. The system is initially in the ground state. We observe that the closer the qubit is to the center mode~$\cent$ in frequency, the worse the gate fidelity is. This occurs because the hybridization between the single-excitation states of qubit~$\qub_1$ and the center mode~$\cent$ increases as they become closer to resonance, resulting in a non-zero matrix element driving the transition to the excited center-mode state (see App.~\ref{app:hybrcrosstalk}). On the other hand, running parallel single-qubit gates on both~$\qub_1$ and~$\qub_2$ does not noticeably alter the gate fidelity, suggesting that for the chosen parameters the hybridization is suppressed between the computational qubits, and only occurs between the qubits and the coupling modes. An exception is if the qubits~$\qub_1$ and~$\qub_2$ are very close to resonance, where there is a sharp increase in error. At this frequency the two computational modes are strongly hybridized and, consequently, the local qubit drive couples strongly to both computational qubits. Running the gates simultaneously removes the error because both computational modes are driven equally. However, if one would perform single-qubit gates with different~$\theta$ and~$\phi$ angles, the hybridization would again disturb the gate close to the~$\qub_1-\qub_2$ resonance. Nevertheless, since the infidelity peak is narrow, this suggests that there exists a large frequency window, roughly from~$\SIrange{4.9}{5.0}{\giga\hertz}$ for the studied parameters, where the qubits can be placed so that they are protected from the hybridization crosstalk.

The above discussion should be contrasted with the case in which we make a single-qubit gate to the center mode~$\cent$, instead of~$\qub_2$. In such a case, the error is not so large for small detunings. The single-excitation states are strongly hybridized, but the two local gate drives compensate for that (again performing gates with different angles would increase the error). However, as the idling frequency of~$\qub_1$ is increased towards the resonance between the two-excitation states~$\ket{200}$ and~$\ket{101}$, the error again increases. Finally, we show a situation where we only perform single-qubit gate on~$\qub_1$, but the center mode is initially excited. For small detuning the error behaves as in the undisturbed case. As the detuning approaches~$\ket{200}-\ket{101}$ resonance, the behaviour changes and resembles the case where also the center mode is driven. 

Since the central mode is considered a non-computational element in our device, it is ideally only temporarily excited during the CZ gate. Assuming that one does not run single-qubit gates on neighbouring qubits during the CZ gate operation, the last two scenarios should not occur on our device. They can occur, however, on a conventional square lattice systems, where the data qubits are typically separated by only a single coupling mode. Thus, our device should be better protected against the hybridization crosstalk than the conventional square lattice. Based on Fig.~\ref{fig:spectators}, the center mode gains a small population ($\sim 10^{-5}$) after each gate operation which can accumulate in longer algorithms with several repeated CZ operations within the unit cell, similar to the Landau--Zener effect~\cite{Landau32,Zener32,Stuckelberg32, Majorana1932}. As Fig.~\ref{fig:sqg} then shows, this population accumulation can start to affect the single-qubit-gate performances, suggesting that the capability to reset the center mode has to be included in the final design.

The optimized driving frequencies and pulse amplitudes, respectively, for the qubits and the center mode are shown in~Fig.~\ref{fig:sqg}~(b) and~(c).

\subsection{Hybridization in the computational states}
In order to understand the origin of the reduced hybridization crosstalk in our device, and to illustrate the (de)localization in the computational basis, we compute all the computational states by solving the eigenproblem for the case $N=6$ of the unit-cell Hamiltonian shown in Eq.~\eqref{eq:ham1}. For each computational state, we compute the expectation values of the number operators~$\nop_\ell$ for each local qubit, and the combined coupler mode consisting of the center mode and the tunable couplers. Also for each computational state, we compute the expectation value of the total number operator~$\nop_{\rm tot} = \sum_{\ell}\nop_\ell$, round the obtained value to the nearest integer, and form groups of computational states with the same total number of excitations. Finally, we overlay the local-qubit occupations for each computational state on a schematic of the honeycomb unit cell and illustrate the data in Fig.~\ref{fig:compstates}. 

We observe that the hybridization between the qubits is very small in all computational states, and the strongest hybridization is with the coupler modes, which increases with the number of excitations. This suggests that for the chosen parameters the computational states of the system are well protected from the hybridization crosstalk between the qubits, as also seen in the single-qubit-gate simulations in Fig.~\ref{fig:sqg}. We also note that the ground state contains some local excitations, since the counter rotating terms couple the states with the same parity. More thorough analysis is given in App.~\ref{app:ipr}.

\section{Impact of decoherence}\label{sec:disspation}
\begin{figure}
    \centering
    \includegraphics[width=1.0\linewidth]{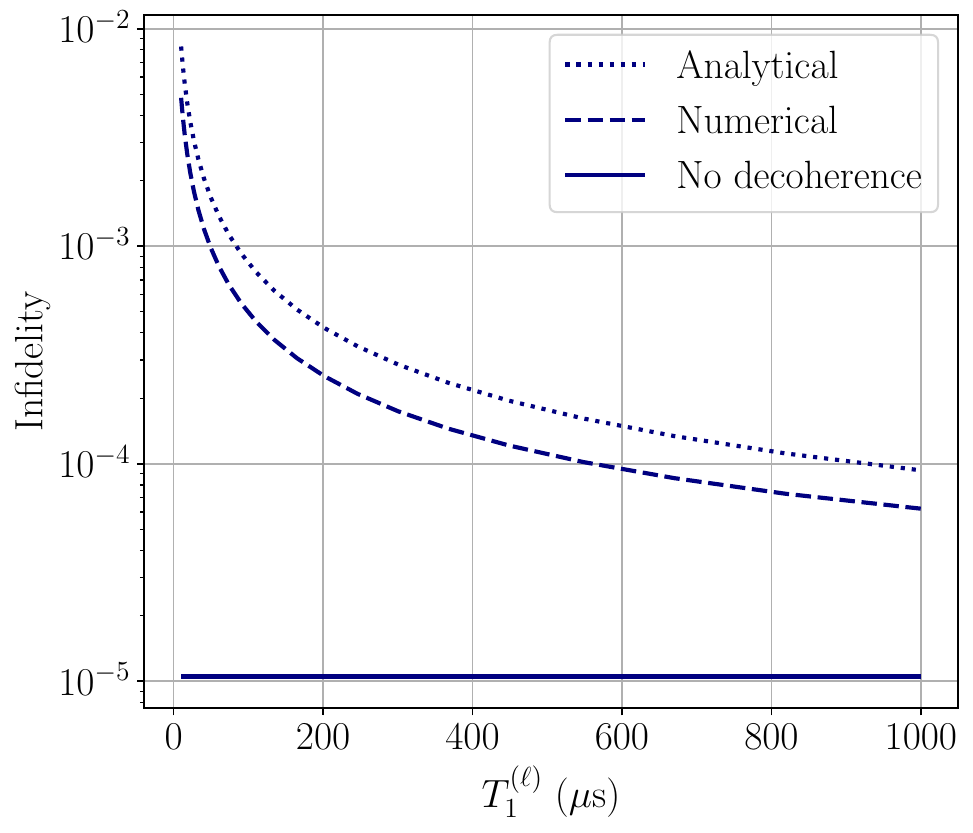}
    \caption{Effect of relaxation and dephasing on the average gate fidelity for~$N=2$ system with gate the duration~$\tau=\SI{60.0}{\nano\second}$. Analytical result (dotted) in Eq.~\eqref{eq:foe_sw} slightly overestimates the simulated average gate infidelity (dashed). The coherent error (solid) has been added to the analytical result, so that both errors converge to the same value as~$T_1^{(\ell)}$ increases. We assume that~$T_2^{*(\ell)} = 2T_1^{(\ell)}$, and the values are the same for all elements, including the couplers. We have used~$\sigma_{\tc}=\SI{3.3}{\nano\second}$.}
    \label{fig:decoh}
\end{figure}
So far we have only considered errors to the ideal gate operation that arise from the coherent unitary dynamics of the system. In reality, the gate fidelities are also limited by relaxation and dephasing of the devices. In the following, we refer to the errors arising from the unitary dynamics as the coherent errors, including leakage and spectator errors, and those resulting from an interaction with some noisy bath as incoherent errors. In App.~\ref{sec:AVGF}, we present Eq.~\eqref{eq:foe2} for calculating the average gate fidelity in an~$N$-qubit system with both coherent and incoherent errors. We apply the equation to the effective model for a system consisting of two computational qubits~$\qub_1$ and~$\qub_2$ and the center mode~$\cent$, described in Eqs.~\eqref{eq:hamiltonians} and~\eqref{eq:propagators}. We model the incoherent dynamics with Lindblad master equation for the density operator $\hat \rho$ of the system, which is given as
\begin{align}
    &\frac{\D\dens}{\D t} = -\frac{\ii }{\hbar}[\hop_{\rm SW}, \dens] + \sum_{\ell=\qub_1,\qub_2,\cent}\gamma_\ell(\bar n_\ell + 1)\mathcal{D}\left[\aop_\ell,\dens\right]
    \nonumber
    \\
    &+ \sum_{\ell=\qub_1,\qub_2,\cent}\gamma_\ell \bar n_\ell\mathcal{D}\left[\aop_\ell^\dag,\dens\right]
    + \sum_{\ell=\qub_1,\qub_2,\cent}2\kappa_\ell^\phi\mathcal{D}\left[\nop_\ell,\dens\right],
\end{align}
where $\mathcal{D}\left[\lop_k,\dens\right] = \lop_k\dens\lop_k^\dag -\frac{1}{2}\left\{\lop_k^\dag\lop_k,\dens\right\}$, $\gamma_\ell$ and $\kappa_\ell^\phi$ are the relaxation and dephasing rates of the mode $\ell$, and the average number of quanta in the environment is given by the Bose--Einstein distribution as
\begin{equation}
    \bar n_\ell = \frac{1}{e^{\beta_\ell\hbar\omega_\ell} - 1},
\end{equation}
where~$\beta_\ell = 1/(k_{\rm B}T_\ell)$ and $T_\ell$ is the temperature of the bath of the mode $\ell$. Using Eq.~\eqref{eq:foe2} for the CZ gate implemented with the unitaries in Eq.~\eqref{eq:propagators}, we obtain the average CZ-gate fidelity
\begin{align}
    \bar{\mathcal{F}}(\tau) &=
    1-\tau\Bigg[\frac{589}{1280}\gamma_{\qub_1}(\bar n_{\qub_1}+1)
    +\frac{7}{32}\gamma_{\qub_2}(\bar n_{\qub_2}+1) 
    \nonumber \\
    &+\frac{1}{8}\gamma_{\cent}(\bar n_{\cent}+1)
    +\frac{1869}{1280}\gamma_{\qub_1}\bar n_{\qub_1}
    +\frac{39}{32}\gamma_{\qub_2}\bar n_{\qub_2}
    \nonumber\\
    &+\frac{9}{8}\gamma_{\cent}\bar n_{\cent}
    +\frac{6373}{10240}\kappa_{\qub_1}^\phi 
    +\frac{633}{2048}\kappa_{\qub_2}^\phi 
    +\frac{141}{640}\kappa_{\cent}^\phi 
    \Bigg],
\end{align}
where~$\tau$ is the duration of the gate. In the zero-temperature limit, the bath occupations are~$\bar n_\ell = 0$, and we obtain
\begin{align}
    \bar{\mathcal{F}}(\tau)
    &=
    1-\frac{589}{1280}\frac{\tau}{T_1^{(\qub_1)}}
    -\frac{7}{32}\frac{\tau}{T_1^{(\qub_2)}}
    -\frac{1}{8}\frac{\tau}{T_1^{(\cent)}}
    \nonumber\\
    &-\frac{6373}{10240}\frac{\tau}{T_2^{*(\qub_1)}}
    -\frac{633}{2048}\frac{\tau}{T_2^{*(\qub_2)}}
    -\frac{141}{640}\frac{\tau}{T_2^{*(\cent)}},\label{eq:foe_sw}
\end{align}
where~$T_1^{(\ell)} = 1/\gamma_\ell$ and~$T_2^{*(\ell)}=1/\kappa_\ell^\phi$ are the relaxation and pure dephasing times of the mode~$\ell$. Comparison with the corresponding formula for the conventional CZ gate presented in Ref.~\cite{abad2025} shows similar performance.

In Fig.~\ref{fig:decoh} we show the analytical average gate infidelity calculated from Eq.~\eqref{eq:foe_sw} as a function of relaxation time~$T_1^{(\ell)}$ and compare it against a numerical simulation of the full system containing also tunable couplers. The gate duration is again~$\SI{60.0}{\nano\second}$. We assume the relaxation and pure dephasing times to be the same for all elements, including the couplers~$\tc_1$ and~$\tc_2$, and set~$T_2^{*(\ell)}=2T_1^{(\ell)}$. The data in Fig.~\ref{fig:decoh} shows that the analytic formula slightly overestimates the error. We emphasize that even though the numerical model includes the couplers whereas the analytic model does not, the initial coupler population is zero also in the simulation and during the gate operation the coupler modes gain temporarily only a small occupation compared to those of the computational and center modes. Overall, our simple analytic model gives surprisingly good estimation for the incoherent error of the gate, and can thus be used in order-of-magnitude estimates of gate fidelities in devices whose operation is limited by incoherent dynamics.

We emphasize that the dissipation model studied here relies on some crude approximations. For example, in reality the pure dephasing in superconducting circuits is typically dominated by~$1/f$ noise, instead of the white noise~\cite{Marxer2023} assumed by the Lindblad equation. Moreover, the time dependence in the Hamiltonian should be reflected into the dissipation rates~\cite{Tuorila2017}. However, since the main focus of this work is in the coherent dynamics, we leave a more accurate description of dissipation in this system for a later study.

\section{Conclusions}\label{sec:conclusions}
In this work we studied theoretically the single- and two-qubit gate operation in the unit cell of a quantum processor with the honeycomb lattice topology. The tunable coupling between the peripheral qubits of the honeycomb unit cell was created through multi-mode coupling structure in which each qubit is coupled to a center mode through a dedicated tunable coupler. We proposed a novel pulse schedule that can be applied for CZ-gate operation between any computational qubit pair within the unit cell. We optimized the CZ-gate performance numerically and demonstrated the gate operation using a pulse schedule consisting of four different flux pulses applied on the qubits and their respective couplers simultaneously. According to the presented simulation data, the studied CZ gates can be operated relatively fast and with high accuracy, even in the presence of spectating qubits in the unit cell, reaching fidelities above~99.99\%. This increases the connectivity in quantum processors built using the underlying honeycomb unit-cell structure as compared with square-lattice topologies conventionally used in the quantum-processor design, without sacrificing the parallel execution of the CZ gates.

Similar CZ gate scheme could work also on a regular square lattice, where one of the data qubits could take the role of the additional coupling mode, enabling direct CZ gates between next-nearest neighbour qubits, provided that the coupling qubit is initially in the ground state. In general this might be too strict requirement, but it could be possible to design the algorithms in such a way that one could exploit our gate scheme also in other topologies.

We also demonstrated that the single-qubit gates in this device are better protected against hybridization crosstalk than in a more conventional square lattice, due to the additional coupling modes that filter the unprompted interactions between the qubits. Thus there is more freedom in choosing the parking frequency distribution for the computational qubits in the unit cell. For simplicity, we concentrated on pulses with simple analytical expressions, and leave pulse-shape optimization for a later study. Also, we provided an analytical formula that can be used to estimate the decoherence-limited gate fidelities if the relaxation and coherence times of the system are known. However, we left a more detailed study of incoherent errors in the device for future work, and focused our studies on the coherent-error mechanisms that determine the fundamental limits for the accuracy and speed of the quantum gates. 

We conclude that when compared to conventional square lattice solutions, our novel quantum-processor architecture results in an enhanced local connectivity while allowing a maximum level of parallelism. Such features are crucial in the implementation of novel high-weight qLDPC codes, such as tile codes.

\section*{Acknowledgements}
We thank Otto Salmenkivi, Eelis Takala and~$\tilde{\mathfrak{S}}$ for useful discussions and technical support. We also thank Juha Hassel, Eric Hyyppä, Alpo Välimaa and Antti Vepsäläinen for providing comments for the manuscript. 
This work was supported by Business Finland through project CfoQ~(787/31/2025) and the German Federal Ministry for Research, Technology and Space via the grant project Ad~Astra~(FKZ: 13N17243).

\appendix
\section{Effect of the Hamiltonian parameters}\label{app:ham_par}
\begin{figure*}
    \centering
    \includegraphics[width=1.0\linewidth]{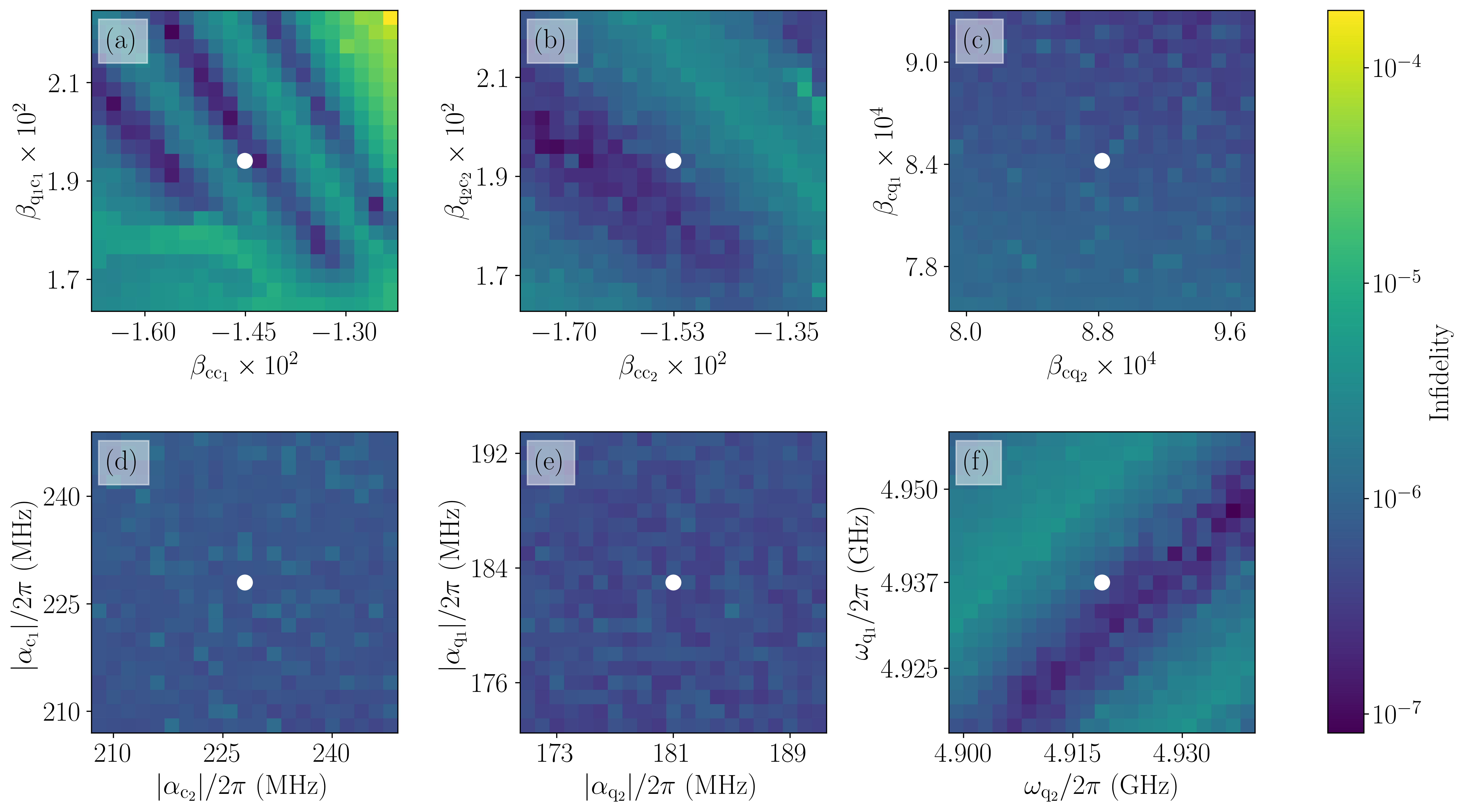}
    \caption{Optimized CZ-gate infidelity as a function of different parameters of the Hamiltonian in Eq.~\eqref{eq:minimal_hamiltonian}. White dots denote the parameters in Tab.~\ref{tab:parameters}. For each data point, we optimize the four pulse amplitudes and the coupler pulse rise time~$\sigma_{\tc}$. We show the gate infidelity as a function of (a)~couplings of the qubit~$\qub_1$ and the center mode~$\cent$ to their shared coupler~$\tc_{1}$; (b)~couplings of the qubit~$\qub_2$ and the center mode~$\cent$ to their shared coupler~$\tc_{2}$; (c)~direct couplings between the qubits and the center mode; (d)~anharmonicities of the tunable couplers; (e)~anharmonicities of the qubits; and (f)~idling frequencies of the qubits. In all cases the qubit~$\qub_2$ is pulsed close to resonance with the center mode during the gate.}
    \label{fig:robustness}
\end{figure*}
In this paper we have used parameters in Tab.~\ref{tab:parameters}, which are close to the values available in a device currently under development~\cite{iqm}. Here we show that the gate performs similarly in a wide range of parameters. For simplicity, we consider a system without spectator qubits, and we also ignore dissipation and dephasing. We choose a pair of parameters (for example idling frequencies of the qubits~$\omega_{\qub_1}$ and~$\omega_{\qub_2}$), and sweep them over a range of values, keeping all the other parameters the same as in Tab.~\ref{tab:parameters}. For each parameter value, we perform the gate optimization as described in Sec.~\ref{sec:gate_optim} as a function of the coupler pulse rise time~$\sigma_{\rm c}$, and choose the rise time which gives the best fidelity. Results are shown in Fig.~\ref{fig:robustness}. In particular, we observe that the values of the qubit and coupler anharmonicities, as well as the direct qubit-center element couplings have practically no effect on the gate fidelity. The biggest impact seems to arise from the couplings to the coupler between the higher frequency qubit and the center mode, where large asymmetry increases the error. Overall the gate infidelity stays below~$10^{-4}$ for most of the parameters, and thus the gate scheme is extremely robust in the system parameters.

\section{Effective Hamiltonian with Schrieffer--Wolff transformation}\label{sec:SF}

Here, we provide the derivation of the effective model in Eq.~\eqref{eq:SF_ham} for a system consisting of two qubits coupled to a center mode through dedicated tunable couplers. We develop a Schrieffer--Wolff type of diagonalization that leads into decoupling of the couplers from the dynamics of the qubits and the center mode.

The Hamiltonian in Eq.~\eqref{eq:minimal_hamiltonian} can be written as $\hop = \hop_0 + \vop$, where
\begin{align}
    \hop_0/\hbar =& \sum_{\ell=\qub_1,\qub_2,\cent,\tc_1,\tc_2}\left[\omega_\ell \nop_\ell + \frac{\alpha_\ell}{2}\nop_\ell(\nop_\ell-1)\right]
    \nonumber\\
    &- \sum_{j=1}^2 g_{\cent\qub_j}\left(\aop_\cent^\dag-\aop_\cent\right)\left(\aop_{\qub_j}^\dag-\aop_{\qub_j}\right) 
    \nonumber\\
    \vop/\hbar =&- \sum_{\ell=\qub_1,\qub_2,\cent}\sum_{j=1}^2g_{\ell\tc_j}\left(\aop_{\ell}^\dag-\aop_{\ell}\right)\left(\aop_{\tc_j}^\dag-\aop_{\tc_j}\right).\label{eq:minimal_hamiltonian_app}
\end{align}
The couplers are approximately decoupled to first order in the coupling rates $g_{\ell \tc_j}$ with a unitary transformation $\hat U = e^{-\sop}$. By expanding the transformed Hamiltonian using the Baker--Campbell--Hausdorff lemma as
\begin{equation}\label{eq:sf1}
    \hop_{S} = e^{-\sop}\hop e^{\sop} 
    = \hop + [\hat S,\hop] 
    +\frac{1}{2}[\hat S, [\hat S, \hop]] +\dots,
\end{equation}
the first order couplings can be eliminated if
\begin{equation}\label{eq:sf2}
    \hat V + [\hat S, \hop_0] = 0.
\end{equation}
With this choice, the transformed Hamiltonian in Eq.~\eqref{eq:sf1} is written up to second order as
\begin{equation}\label{eq:schieffer_wolff}
    \hop_{S} \approx
    \hop_0 + \frac{1}{2}[\hat S,\hat V].
\end{equation}
For the Hamiltonian defined in Eq.~\eqref{eq:minimal_hamiltonian_app}, the transformation fulfilling Eq.~\eqref{eq:sf2} is given as~\cite{Yan2018}
\begin{align}
    \hat S
    &=
    \frac{g_{\qub_1\tc_1}}{\Delta_{\qub_1\tc_1}}
    \left(\aop_{\qub_1}^\dag\aop_{\tc_1}
    -\aop_{\qub_1}\aop_{\tc_1}^\dag\right)
    -\frac{g_{\qub_1\tc_1}}{\Sigma_{\qub_1\tc_1}}
    \left(\aop_{\qub_1}^\dag\aop_{\tc_1}^\dag
    -\aop_{\qub_1}\aop_{\tc_1}\right)\nonumber
    \\
    &+
    \frac{g_{\qub_2\tc_2}}{\Delta_{\qub_2\tc_2}}
    \left(\aop_{\qub_2}^\dag\aop_{\tc_2}
    -\aop_{\qub_2}\aop_{\tc_2}^\dag\right)
    -\frac{g_{\qub_2\tc_2}}{\Sigma_{\qub_2\tc_2}}
    \left(\aop_{\qub_2}^\dag\aop_{\tc_2}^\dag
    -\aop_{\qub_2}\aop_{\tc_2}\right)\nonumber\\
    &+
    \frac{g_{\cent\tc_1}}{\Delta_{\cent \tc_1}}
    \left(\aop_{\cent}^\dag\aop_{\tc_1}
    -\aop_{\cent}\aop_{\tc_1}^\dag\right)
    -\frac{g_{\cent\tc_1}}{\Sigma_{\cent \tc_1}}
    \left(\aop_{\cent}^\dag\aop_{\tc_1}^\dag
    -\aop_{\cent}\aop_{\tc_1}\right)\nonumber\\
    &+
    \frac{g_{\cent\tc_2}}{\Delta_{\cent \tc_2}}
    \left(\aop_{\cent}^\dag\aop_{\tc_2}
    -\aop_{\cent}\aop_{\tc_2}^\dag\right)
    -\frac{g_{\cent\tc_2}}{\Sigma_{\cent \tc_2}}
    \left(\aop_{\cent}^\dag\aop_{\tc_2}^\dag
    -\aop_{\cent}\aop_{\tc_2}\right),
\end{align}
where we have defined the detuning~$\Delta_{\ell\ell'} \equiv \omega_{\ell} - \omega_{\ell'}$ and the sum~$\Sigma_{\ell\ell'} \equiv \omega_{\ell} + \omega_{\ell'}$ of the frequencies. Computing the commutator in Eq.~\eqref{eq:schieffer_wolff} results in the Hamiltonian in Eq.~\eqref{eq:SF_ham}, with parameters defined as
\begin{align}
    \widetilde\omega_{\qub_j}
    &\equiv
    \omega_{\qub_j} +g_{\qub_j\tc_j}^2\left(
    \frac{1}{\Delta_{\qub_j\tc_j}}-\frac{1}{\Sigma_{\qub_j\tc_j}}
    \right),
    \nonumber
    \\
    \widetilde\omega_{\cent}
    &\equiv
    \omega_{\cent}
    + g_{\cent\tc_1}^2\left(
    \frac{1}{\Delta_{\cent\tc_1}}
    -\frac{1}{\Sigma_{\cent\tc_1}}
    \right)
    + g_{\cent\tc_2}^2\left(
    \frac{1}{\Delta_{\cent\tc_2}}
    -\frac{1}{\Sigma_{\cent\tc_2}}
    \right),
    \nonumber
    \\
    \widetilde\omega_{\tc_j}
    &\equiv\omega_{\tc_j}
    -g_{\qub_j\tc_j}^2\left(
    \frac{1}{\Delta_{\qub_j\tc_j}}+\frac{1}{\Sigma_{\qub_j\tc_j}}
    \right)
    -g_{\cent\tc_j}^2\left(
    \frac{1}{\Delta_{\cent\tc_j}}
    +\frac{1}{\Sigma_{\cent\tc_j}}
    \right),
    \nonumber
    \\
    \widetilde g_{\cent\qub_j} 
    &\equiv
    g_{\cent\qub_j} + 
    \frac{g_{\qub_j\tc_j}g_{\cent\tc_j}}{2}
    \left(
    \frac{1}{\Delta_{\qub_j\tc_j}} +
    \frac{1}{\Delta_{\cent\tc_j}}-
    \frac{1}{\Sigma_{\qub_j\tc_j}}-
    \frac{1}{\Sigma_{\cent\tc_j}}
    \right),
    \nonumber
    \\
    \widetilde g_{\tc_1\tc_2}
    &\equiv
    \frac{g_{\cent\tc_1}g_{\cent\tc_2}}{2}
    \left(
    \frac{1}{\Delta_{\cent\tc_1}} +
    \frac{1}{\Delta_{\cent\tc_2}} +
    \frac{1}{\Sigma_{\cent\tc_1}}+
    \frac{1}{\Sigma_{\cent\tc_2}}
    \right),
\end{align}
where~$j=1,2$. For simplicity we have neglected the effect of the transformation on the anharmonicities. 

\section{Jacobian diagonalization in the single-excitation manifold}\label{sec:Jacobi}
\begin{figure}
    \centering
    \includegraphics[width=1.0\linewidth]{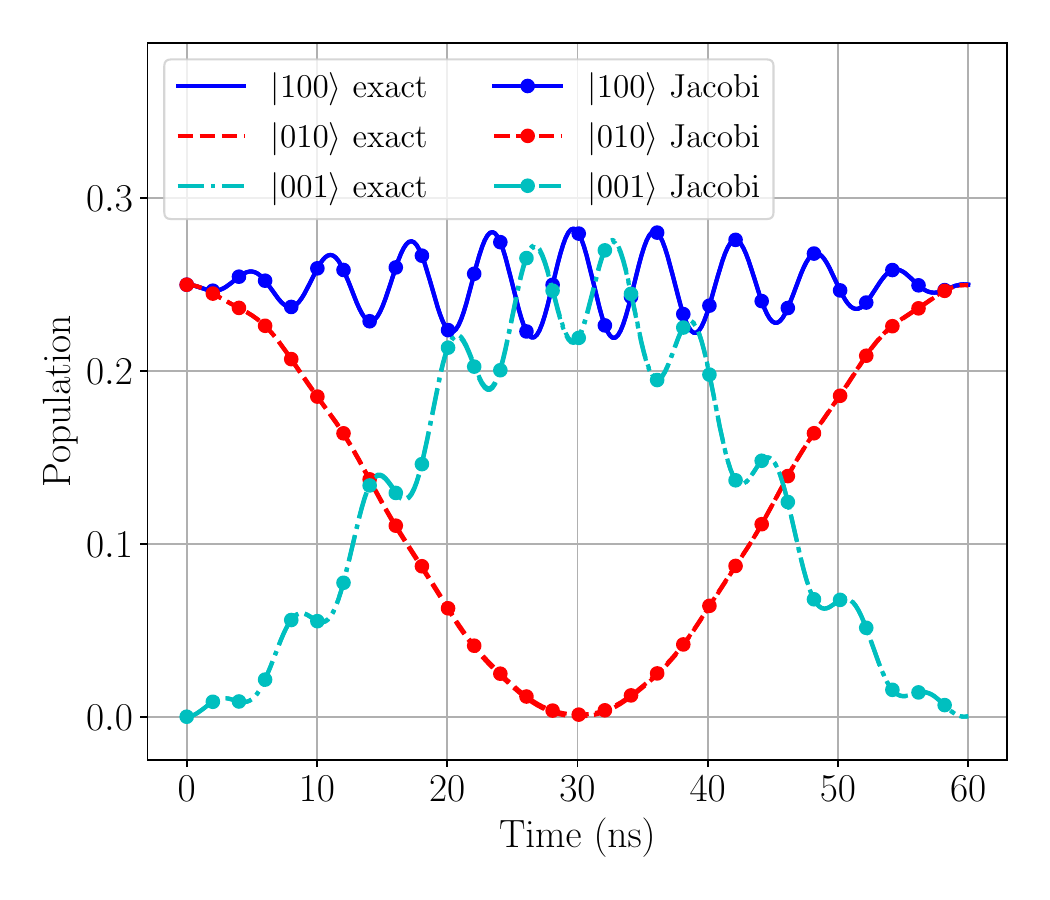}
    \caption{Comparison of exactly solved single-excitation dynamics and the result obtained with Jacobian diagonalization using Eq.~\eqref{eq:approximative_propagator} (dots as markers). The Jacobian diagonalization gives a good estimate for the actual dynamics.}
    \label{fig:jacobi_accuracy}
\end{figure}
In Sec.~\ref{sec:gate} we assumed that in the single-excitation manifold the coupling between the off resonant state~$\ket{100}$ and the resonant states~$\ket{010}$ and~$\ket{001}$ can be neglected in Eq.~\eqref{eq:hamiltonians}. This is naturally not a very good approximation since the coupling strengths~$\widetilde g_{\cent\qub_1}$ and~$\widetilde g_{\cent\qub_2}$ are of the same order of magnitude. This is visible in the data shown in Fig.~\ref{fig:populations} for the full numerical model. The population in the non-resonant state oscillates in time and, thus, can potentially lead to errors in the gate operation if not calibrated appropriately. Therefore, it is worthwhile to analyze the simple model also taking into account the coupling to the off-resonant state~$\ket{100}$. 

Here, we attempt a more accurate approach using the Jacobian diagonalization~\cite{Jacobi1846, Tuorila2017} for the single-excitation part of the Hamiltonian in Eq.~\eqref{eq:hamiltonians}. The Hamiltonian is written in the basis~$\{\ket{001},\,\ket{010},\,\ket{100}\}$ with indexing~$\ket{n_{\qub_1} n_{\qub_2} n_{\cent}}$, and is of the form
\begin{equation}\label{eq:jacobi_exact_ham}
    \hop \equalhat \hbar 
    \begin{pmatrix}
        \omega_b & g_2 & g_1\\
        g_2 & \omega_b & 0\\
        g_1 & 0 & \omega_a
    \end{pmatrix},
\end{equation}
where we have simplified the notation by defining~$\omega_a=\widetilde\omega_{\qub_1}$,~$\omega_b=\widetilde\omega_{\qub_2} = \widetilde\omega_{\cent}$,~$g_i = \widetilde g_{\cent\qub_i}$ and assumed that~$\omega_a>\omega_b$. Jacobian diagonalization is an iterative procedure which is repeated until the Hamiltonian is diagonal up to the desired accuracy. For each iteration step, we choose the pair of states with the largest coupling and diagonalize the corresponding $2\times 2$ block of the full $3\times 3$ matrix. In the first step we diagonalize the resonant subspace~$\{\ket{001},\,\ket{010}\}$ and obtain
\begin{align}
    \dop_1 = 
    \vop_0^\dag \hop\vop_0
    \equalhat
    \hbar
    \begin{pmatrix}
        \lambda_-^{(0)} & 0 & -\frac{g_1}{\sqrt{2}}\\
        0 & \lambda_+^{(0)} & \frac{g_1}{\sqrt{2}}\\
        -\frac{g_1}{\sqrt{2}} & \frac{g_1}{\sqrt{2}} & \omega_a
    \end{pmatrix},
\end{align}
where~$\lambda_{\pm}^{(0)} \equiv \omega_b\pm g_2$, and the transformation unitary is given as
\begin{equation}
    \vop_0 \equalhat
    \frac{1}{\sqrt{2}}
    \begin{pmatrix}
        -1 & 1 & 0\\
        1 & 1 & 0\\
        0 & 0 & \sqrt{2}
    \end{pmatrix}
    =
    \begin{pmatrix}
        \ket{\lambda_-^{(0)}} & \ket{\lambda_+^{(0)}} & \ket{100}
    \end{pmatrix}
\end{equation}
The state~$\ket{100}$ couples to both states~$\ket{\lambda_\pm^{(0)}}$ with an equal magnitude, but is closer in frequency to~$\ket{\lambda_+^{(0)}}$. Thus, for the next step we diagonalize the subspace~$\{\ket{\lambda_+^{(0)}},\,\ket{100}\}$ and obtain
\begin{align}
    \dop_2 = \vop_1^\dag\dop_1\vop_1
    \equalhat
    \hbar
    \begin{pmatrix}
        \lambda_-^{(0)} & \frac{g_1v_{12}}{\sqrt{2}} & -\frac{g_1v_{11}}{\sqrt{2}}\\
         \frac{g_1v_{12}}{\sqrt{2}} & \lambda_-^{(1)} & 0\\
         -\frac{g_1v_{11}}{\sqrt{2}} & 0 & \lambda_+^{(1)}
    \end{pmatrix},
\end{align}
with the transformation unitary
\begin{equation}
    \vop_1 \equalhat
    \begin{pmatrix}
        1 & 0 & 0\\
        0 & v_{11} & v_{12}\\
        0 & -v_{12} & v_{11}
    \end{pmatrix}.
\end{equation}
The new eigenvalues are
\begin{equation}
    \lambda_\pm^{(1)} \equiv \frac{\omega_a +\lambda_+^{(0)}}{2}\pm\frac{1}{2}\sqrt{\left(\omega_a - \lambda_+^{(0)}\right)^2 + 2g_1^2},
\end{equation}
and the weights in the eigenstates are defined as
\begin{align}
    v_{11} &= \frac{\frac{g_1}{\sqrt{2}}}{\sqrt{\frac{g_1^2}{2} 
    + \left(\lambda_-^{(1)} - \lambda_+^{(0)}\right)^2}},
    \nonumber
    \\
    v_{12} &= \frac{\frac{g_1}{\sqrt{2}}}{\sqrt{\frac{g_1^2}{2} 
    + \left(\lambda_+^{(1)} - \lambda_+^{(0)}\right)^2}}.
\end{align}
The states~$\ket{100}$ and~$\ket{\lambda_+^{(0)}}$ are off resonant, and thus their coupling only slightly shifts the energy levels, suggesting that~$\lambda_-^{(1)}\approx\lambda_+^{(0)}$ and~$\lambda_+^{(1)}\approx\omega_a$. We therefore have that~$\lambda_-^{(1)} - \lambda_+^{(0)}< \lambda_+^{(1)} - \lambda_+^{(0)}$, which leads to~$|v_{12}| < |v_{11}|$. We set~$\frac{g_1v_{12}}{\sqrt{2}}\approx 0$, and obtain the approximative Hamiltonian
\begin{align}
    \dop
    \equalhat
    \hbar
    \begin{pmatrix}
        \lambda_-^{(0)} & 0 & -\frac{g_1v_{11}}{\sqrt{2}}\\
         0 & \lambda_-^{(1)} & 0\\
         -\frac{g_1v_{11}}{\sqrt{2}} & 0 & \lambda_+^{(1)}
    \end{pmatrix},
\end{align}
from which we calculate the approximative propagator
\begin{equation}\label{eq:approximative_propagator_D}
    \uop_D(t) \equalhat
    \begin{pmatrix}
        r(t)e^{-\ii \left[\bar\lambda t - \varphi(t)\right]} & 0 & \ii c(t)e^{-\ii \bar\lambda t}  \\
        0 & e^{-\ii \lambda_-^{(1)}t} & 0\\
        \ii c(t)e^{-\ii \bar\lambda t} & 0 & r(t)e^{-\ii \left[\bar\lambda t + \varphi(t)\right]}
    \end{pmatrix},
\end{equation}
where we have defined
\begin{align}
    r(t) &\equiv \sqrt{\cos^2\left(\frac{\varpi t}{2}\right) 
    + \frac{\Delta^2}{\varpi^2}\sin^2\left(\frac{\varpi t}{2}\right)},
    \nonumber\\
    c(t) &\equiv \frac{2g_1v_{11}}{\sqrt{2}\varpi}\sin\left(\frac{\varpi t}{2}\right),
    \nonumber\\
    \varphi(t) &\equiv \arctan\left[\frac{\Delta}{\varpi}\tan\left(\frac{\varpi t}{2}\right)\right],
\end{align}
and
\begin{align}
    \bar\lambda &\equiv \frac{\lambda_+^{(1)} + \lambda_-^{(0)}}{2},\quad
    \Delta \equiv \lambda_+^{(1)} - \lambda_-^{(0)},\nonumber\\
    \varpi&\equiv \sqrt{\Delta^2 + 2g_1^2v_{11}^2}.
\end{align}
The Jacobian diagonalization up to this point has been
\begin{equation}
    \dop_2 = \vop_1^\dag\dop_1\vop_1 = \vop_1^\dag\vop_0^\dag\hop\vop_0\vop_1
    = \vop^\dag \hop\vop,
\end{equation}
where
\begin{equation}
    \vop \equiv \vop_0\vop_1
    \equalhat
    \frac{1}{\sqrt{2}}
    \begin{pmatrix}
        -1 & v_{11} & v_{12}\\
        1 & v_{11} & v_{12}\\
        0 & -\sqrt{2}v_{12} & \sqrt{2}v_{11}
    \end{pmatrix}.
\end{equation}
In the original basis the propagator in Eq.~\eqref{eq:approximative_propagator_D} is then
\begin{equation}
    \uop(t) = \vop\uop_D(t)\vop^\dag,
\end{equation}
which has the matrix elements
\begin{widetext}
    \begin{align}
        \bra{001}\uop(t)\ket{001} 
        &=
        \frac{v_{11}^{2} e^{- \ii  \lambda_-^{(1)} t}}{2} + \frac{v_{12}^{2} r{\left(t \right)} e^{- \ii  \left[\bar\lambda t + \varphi{\left(t \right)}\right]}}{2} - \ii  v_{12} c{\left(t \right)} e^{- \ii  \bar\lambda t} + \frac{r{\left(t \right)} e^{- \ii \left[ \bar\lambda t - \varphi{\left(t \right)}\right]}}{2}
        \nonumber
        \\
        \bra{001}\uop(t)\ket{010} 
        &=
        \frac{v_{11}^{2} e^{- \ii  \lambda_-^{(1)} t}}{2} + \frac{v_{12}^{2} r{\left(t \right)} e^{- \ii  \left[\bar\lambda t +  \varphi{\left(t \right)}\right]}}{2} - \frac{r{\left(t \right)} e^{- \ii  \left[\bar\lambda t -  \varphi{\left(t \right)}\right]}}{2}
        \nonumber
        \\
        \bra{001}\uop(t)\ket{100} 
        &=
        \frac{v_{11} v_{12} r{\left(t \right)} e^{- \left[\ii  \bar\lambda t +  \varphi{\left(t \right)}\right]}-
        v_{11} v_{12} e^{- \ii  \lambda_-^{(1)} t} - \ii  v_{11} c{\left(t \right)} e^{- \ii  \bar\lambda t}}{\sqrt{2}}
        \nonumber
        \\
        \bra{010}\uop(t)\ket{001}
        &=\bra{001}\uop(t)\ket{010},
        \nonumber
        \\
        \bra{010}\uop(t)\ket{010}
        &=
        \frac{v_{11}^{2} e^{- \ii  \lambda_-^{(1)} t}}{2} + \frac{v_{12}^{2} r{\left(t \right)} e^{- \ii  \left[\bar\lambda t +  \varphi{\left(t \right)}\right]}}{2} + \ii  v_{12} c{\left(t \right)} e^{- \ii  \bar\lambda t} + \frac{r{\left(t \right)} e^{- \ii  \left[\bar\lambda t -  \varphi{\left(t \right)}\right]}}{2} 
        \nonumber
        \\
        \bra{001}\uop(t)\ket{100}
        &=
        \frac{v_{11} v_{12} r{\left(t \right)} e^{- \left[\ii  \bar\lambda t +  \varphi{\left(t \right)}\right]}
        -v_{11} v_{12} e^{- \ii  \lambda_-^{(1)} t}
        +\ii  v_{11} c{\left(t \right)} e^{- \ii  \bar\lambda t}}{\sqrt{2}}
        \nonumber
        \\
        \bra{100}\uop(t)\ket{001}
        &= \bra{001}\uop(t)\ket{100}
        \nonumber
        \\
        \bra{100}\uop(t)\ket{010}
        &=\bra{010}\uop(t)\ket{100}
        \nonumber
        \\
        \bra{100}\uop(t)\ket{100}
        &=
        v_{11}^{2} r{\left(t \right)} e^{- \ii \left[\bar\lambda t +  \varphi{\left(t \right)}\right]} + v_{12}^{2} e^{- \ii  \lambda_-^{(1)} t}.
        \label{eq:approximative_propagator}
    \end{align}
\end{widetext}
Comparison between the numerically solved dynamics using the Hamiltonian in Eq.~\eqref{eq:jacobi_exact_ham} and the one using the approximative propagator in Eq.~\eqref{eq:approximative_propagator} is shown in Fig.~\ref{fig:jacobi_accuracy}. The propagator obtained with the Jacobian diagonalization produces very accurate estimation for the system dynamics, but the resulting form is not particularly intuitive.

For the gate performance it is essential that all state populations in the single- and two-excitation manifold return to their initial values at the end of the gate. In the two-excitation manifold this is achieved if all relevant states are in resonance. In the single-excitation manifold, on the other hand, the detuned higher-frequency qubit state performs off-resonant Rabi oscillations which interfere with the desired resonant oscillations between the lower frequency qubit and the center mode. This can cause population swapping between the qubits, which hinders the accuracy of the gate. This was indeed observed in Fig.~\ref{fig:spectators} as the main source of error in the populations.

\section{MOVE-CZ-MOVE}\label{app:mczm}
In order to highlight the difference between our novel gate scheme presented in Sec.~\ref{sec:analyt_mod} and the MOVE-CZ-MOVE introduced in Ref.~\onlinecite{Renger2025}, in this section we provide the derivation of the gate unitary for the MOVE-CZ-MOVE. The unitary consists of three successive gate operations. The Hamiltonian describing the dynamics is given by Eq.~\eqref{eq:hamiltonians}, but now the couplings are time-dependent. Switching to a frame rotating with qubit and center element frequencies with the time-dependent unitary transformation ~$\rop(t) = \exp\left[\ii t\sum_{\ell = \qub_1,\qub_2,\cent}\widetilde\omega_{\ell}\nop_\ell\right]$, the Hamiltonian~\eqref{eq:hamiltonians} in the relevant subspaces is written as
\begin{align}
    \hop_1(t) &\equalhat\hbar
    \begin{pmatrix}
        0 & g_2(t) & g_1(t) e^{-\ii \Delta t} \\
        g_2(t) & 0 & 0 \\
        g_1(t) e^{\ii \Delta t} & 0 & 0
    \end{pmatrix},
    \nonumber
    \\
    \hop_2(t) &\equalhat\hbar
    \begin{pmatrix}
        0 & 0 & \sqrt{2}g_1(t) \\
        0 & 0 & g_2(t) \\
        \sqrt{2}g_1(t) & g_2(t) & 0
    \end{pmatrix},
\end{align}
where we have simplified the notation by defining~$\omega_a=\widetilde\omega_{\qub_1}$,~$\omega_b=\widetilde\omega_{\qub_2} = \widetilde\omega_{\cent}$,~$g_i = \widetilde g_{\cent\qub_i}$, and we have defined~$\Delta = \omega_a-\omega_b=-\alpha_1$. During the MOVE gates,~$g_1(t)=0$ and~$g_2(t) = g_2$, and during the CZ gate~$g_1(t) = g_1$ and~$g_2=0$. We assume that the couplings are turned on and off instantaneously. 
For the MOVE gate we readily obtain the propagators in the single- and two-excitation manifolds to be of the form
\begin{widetext}
\begin{align}
    \uop_1^{\rm MOVE}(t, t_0) &\equalhat
    \begin{pmatrix}
        \cos[g_2(t-t_0)] & -\ii \sin[g_2(t-t_0)] & 0\\
        -\ii\sin[g_2(t-t_0)] & \cos[g_2(t-t_0)] & 0\\
        0 & 0& 1
    \end{pmatrix},
    \nonumber
    \\
    \uop_2^{\rm MOVE}(t, t_0) &\equalhat
    \begin{pmatrix}
        1 & 0 & 0\\
        0 & \cos[g_2(t-t_0)] & -\ii \sin[g_2(t-t_0)]\\
        0 & -\ii\sin[g_2(t-t_0)] & \cos[g_2(t-t_0)]
    \end{pmatrix}.
\end{align}
Perfect MOVE operation is achieved by choosing~$g_2\tau_{\rm MOVE} = \pi/2$ with~$\tau_{\rm MOVE} = t - t_0$. The CZ gate is slightly more involved due to off-resonant coupling in the single-excitation manifold. We obtain
\begin{align}
    \uop_1^{\rm CZ}(t,t_0) &\equalhat
    \begin{pmatrix}
        r(t-t_0)e^{\ii\varphi(t-t_0)} e^{-\frac{\ii\Delta (t-2t_0)}{2}}&0 & -\ii c(t-t_0)e^{-\frac{\ii\Delta t}{2}}
        \\
        0 & 1 & 0\\
        -\ii c(t-t_0)e^{-\frac{\ii\Delta t}{2}} & 0 
        & r(t-t_0)e^{-\ii\varphi(t-t_0)} e^{\frac{\ii\Delta (t-2t_0)}{2}}
    \end{pmatrix},
    \nonumber
    \\
    \uop_2^{\rm CZ}(t,t_0) &\equalhat
    \begin{pmatrix}
        \cos\left[\sqrt{2}g_1(t-t_0)\right] & 0 & -\ii\sin\left[\sqrt{2}g_1(t-t_0)\right]\\
        0 & 1 & 0\\
        -\ii\sin\left[\sqrt{2}g_1(t-t_0)\right] & 0 & \cos\left[\sqrt{2}g_1(t-t_0)\right]
    \end{pmatrix},
\end{align}
where we have defined
\begin{align}
    r(t) = \sqrt{\cos^2\left(\frac{\Omega t}{2}\right) 
    + \frac{\Delta^2}{\Omega^2}\sin^2\left(\frac{\Omega t}{2}\right)},
    \quad
    \varphi(t) = \arctan\left[\frac{\Delta}{\Omega}\tan\left(\frac{\Omega t}{2}\right)\right],
    \quad
    c(t) =\frac{2 g}{\Omega}\sin\left(\frac{\Omega t}{2}\right),
\end{align}
\end{widetext}
with~$\Omega = \sqrt{\Delta^2+4g^2}$. Choosing~$\sqrt{2}g_1 \tau_{\rm CZ} = \pi$ (now with~$\tau_{\rm CZ} = t - t_0$ and~$t_0 = \tau_{\rm MOVE}$) gives the CZ gate. The off-resonant oscillations in the single-excitation manifold can be synchronized by demanding that~$\cos(\Omega \tau_{\rm CZ}/2) = 1$, which happens if
\begin{equation}
    g_1 = \frac{\pm \Delta}{2\sqrt{8n^2 - 1}},\, n\in \mathbb{N},
\end{equation}
i.e. by choosing the ratio~$g_1/\alpha_1$ suitably. This also guarantees that~$\varphi(\tau_{\rm CZ}) = 0$. We here write the total propagators as
\begin{align}
    \uop_1(t) &\equalhat 
    \begin{pmatrix}
        -1 & 0 & 0\\
        0 & -e^{\frac{\ii\Delta \tau_{\rm MOVE}}{2}} & 0\\
        0 & 0 & e^{-\frac{\ii\Delta\tau_{\rm MOVE}}{2}}
    \end{pmatrix},\nonumber
    \\
    \uop_2(t) &\equalhat
    \begin{pmatrix}
        -1 & 0 & 0 \\
        0 & 1 & 0 \\
        0 & 0 & -1
    \end{pmatrix}.\label{eq:move-cz-move_unitary}
\end{align}
We observe that the propagator acting within the two-excitation manifold differs from the corresponding propagator in Eq.~\eqref{eq:propagators}, whose diagonal entries are all~$+1$. Here, instead, both non-computational states~$\ket{200}$ and~$\ket{101}$ acquire a~$\pi$-phase, while the computational state~$\ket{110}$ does not. In the single-excitation manifold, similarly as in Eq.~\eqref{eq:propagators}, the states~$\ket{001}$ and~$\ket{010}$ acquire a~$\pi$-phase. Moreover, the computational states~$\ket{100}$ and~$\ket{010}$ accumulate additional phases due to off-resonant oscillation [which was ignored for simplicity in the derivation of Eq.~\eqref{eq:propagators}]. However, these phases are opposite and can be removed without affecting the phase of the state~$\ket{110}$. Following similar procedure as in Sec.~\ref{sec:analyt_mod}, we obtain the effective two-qubit unitary as
\begin{equation}
    \uop_{\qub_1\qub_2} = \begin{pmatrix}
        1 & 0 & 0 & 0\\
        0 & -1 & 0 & 0\\
        0 & 0 & 1 & 0\\
        0 & 0 & 0 & 1
    \end{pmatrix},
\end{equation}
which is exactly the same as in Eq.~\eqref{eq:almost_cz}. Thus, even though the underlying dynamics governed by the propagators in Eqs.~\eqref{eq:propagators} and~\eqref{eq:move-cz-move_unitary} are clearly different, the resulting effective gate unitaries are the same.

\section{Average gate fidelity}\label{sec:AVGF}
In our numerical and analytical calculations, we evaluate the accuracy of the studied gate operation using the concept of average gate fidelity.
The average gate fidelity in an~$N$ qubit system between two quantum processes~$\mathcal{G}_1$ and~$\mathcal{G}_2$ is defined as (if at least one of them keeps the system in a pure state)
\begin{equation}
    \bar{\mathcal{F}} = \int \D{\ket{\psi}}\tr[\mathcal{G}_1(\dens_\psi)\mathcal{G}_2(\dens_\psi)],
\end{equation}
where integration is over all~$N$ qubit initial states~$\ket{\psi}$. Writing the density operator in terms of Pauli spin operators for $N$ qubits and employing certain symmetry relations, we transform the integral into a finite sum over all $N$-qubit Pauli-spin operators $\hat f_j$~\cite{CABRERA2007} and obtain
\begin{equation}\label{eq:general}
    \bar{\mathcal{F}} = \frac{1}{d^2}\Bigg\{
    \tr\left[\mathcal{G}_1(\hat I)\mathcal{G}_2(\hat I)\right]
    + \frac{1}{d+1}\sum_{j=1}^{d^2-1}
    \tr\left[\mathcal{G}_1(\hat f_j)\mathcal{G}_2(\hat f_j)\right]
    \Bigg\},
\end{equation}
where $d = 2^N$, $j = (j_1j_2\cdots j_N)_4$ is the labelling of the operators in base 4, and
\begin{equation}
    \hat f_j = \hat \sigma_{j_1}\otimes
    \hat \sigma_{j_2} \otimes \dots \otimes
    \hat \sigma_{j_N},
\end{equation}
are the all $2^{2N}-1$ different combinations of the Pauli matrices $\sigma_{0} = \hat I$, $\sigma_{1} = \hat \sigma_x$, $\sigma_{2} = \hat \sigma_y$ and $\sigma_{3} = \hat \sigma_z$, excluding the one consisting only of identities, i.e. the case with $j=0$.

Equation~\eqref{eq:general} can be directly used for numerical simulations and analytical calculations for simple unitary processes. Including the effects of relaxation and decoherence analytically is slightly more cumbersome. However, assuming that these effects are small, one can apply perturbation theory~\cite{Martinez2012} on Lindblad master equation~\cite{Villegas2016} and obtain some analytical insight about the incoherent effects. Let us consider a master equation
\begin{align}
    \frac{\D{\dens}}{\D t} = -\frac{\ii }{\hbar}[\hop, \dens] +\sum_{k}\Gamma_k\mathcal{D}\left[\lop_k,\dens\right],
\end{align}
where~$\Gamma_k$ are the jump rates corresponding to the jump operators~$\lop_k$ and
\begin{equation}
    \mathcal{D}\left[\lop_k,\dens\right] = \lop_k\dens\lop_k^\dag -\frac{1}{2}\left\{\lop_k^\dag\lop_k,\dens\right\}.
\end{equation}
Using the perturbation theory for the master equation~\cite{Martinez2012}, for time scales much shorter than $1/\Gamma_k$ we obtain up to first order
\begin{align}\label{eq:me_sol}
    \dens(t) &\approx 
    \hat U(t)\dens^{(0)}(0)\hat U^\dag(t)
    \nonumber
    \\
    &+ \sum_k\Gamma_k\int_0^t \D{t'}
    \hat U(t)\mathcal{D}\left[\lop_k(t'), \dens^{(0)}(0)\right]
    \hat U^\dag(t),
\end{align}
where the dynamics in the zeroth order are determined by the unitary evolution,
\begin{equation}
    \hat U(t) = e^{-\frac{\ii }{\hbar}\hop t},\quad
    \dens^{(0)}(t) = 
    \hat U(t)\dens^{(0)}(0)
    \hat U^\dag(t).
\end{equation}
and the time evolved jump operators are
\begin{equation}
    \lop_k(t) = \hat U^\dag(t)\lop_k\hat U(t),
\end{equation}
where for simplicity we have assumed a time-independent Hamiltonian.

We here consider the quantum operation~$\mathcal{G}_1$ in Eq.~\eqref{eq:general} as some ideal process, say CZ gate, and~$\mathcal{G}_2$ is some realistic dynamical process attempting to produce the desired outcome. Due to imperfections in the implementation, it can contain both coherent errors and decoherence, according to Eq.~\eqref{eq:me_sol}. Moreover, we are in particular interested in the CZ gate, so we set~$N=2$ in Eq.~\eqref{eq:general}, and average over all 16 two qubit Pauli operators, and keep all the others in the ground state. We thus write Eq.~\eqref{eq:general} as
\begin{widetext}
    \begin{align}
        \bar{\mathcal{F}} = 
        &\frac{1}{16}
        \Bigg\{
        \tr
        \left[\mathcal{G}\left(\dens_{\hat I,B}\right)
        \mathcal{U}\left(\dens_{\hat I,B}\right)\right]
        +
        \frac{1}{5}\sum_{j=1}^{15}
        \tr\left[\mathcal{G}\left(\dens_{\hat f_j,B}\right)
        \mathcal{U}\left(\dens_{\hat f_j,B}\right)\right]
        \Bigg\}\nonumber\\
        +
        &\frac{1}{16}\sum_k
        \int_0^t \D{t'}\Gamma_k
        \tr\left\{
        \mathcal{G}\left(\dens_{\hat I,B}\right)
        \hat U(t)
        \mathcal{D}\left[\lop_k(t'),\dens_{\hat I,B}
        \right]
        \hat U^\dag(t)
        \right\}\nonumber\\
        +
        &\frac{1}{80}\sum_k\sum_{j=1}^{15}
        \int_0^t \D{t'}\Gamma_k
        \tr\left\{
        \mathcal{G}\left(\dens_{\hat f_j,B}\right)
        \hat U(t)
        \mathcal{D}\left[\lop_k(t'),\dens_{\hat f_j,B}\right]
        \hat U^\dag(t)
        \right\}, \label{eq:foe2}
    \end{align}
\end{widetext}
\iffalse
\begin{widetext}
    \begin{align}
        \bar{\mathcal{F}} = 
        &\frac{1}{16}
        \Bigg\{
        \tr\left(
        \hat P_{\rm 2Q}
        \left[\tr_{B}\mathcal{G}\left(\dens_{\hat I,B}\right)\right]
        \hat P_{\rm 2Q}
        \left[\tr_{B}\mathcal{U}\left(\dens_{\hat I,B}\right)\right]\right)
        +
        \frac{1}{5}\sum_{j=1}^{15}
        \tr\left(
        \hat P_{\rm 2Q}\left[\tr_{B}\mathcal{G}\left(\dens_{\hat f_j,B}\right)\right]
        \hat P_{\rm 2Q}
        \left[\tr_{B}\mathcal{U}\left(\dens_{\hat f_j,B}\right)\right]
        \right)
        \Bigg\}\nonumber\\
        +
        &\frac{1}{16}\sum_k
        \int_0^t \D{t'}\Gamma_k
        \tr\left\{
        \hat P_{\rm 2Q}
        \left[\tr_{B}\mathcal{G}\left(\dens_{\hat I,B}\right)\right]
        \hat P_{\rm 2Q}
        \tr_{B}\left[
        \hat U(t)
        \mathcal{D}\left[\lop_k(t'),\dens_{\hat I,B}\right]
        \hat U^\dag(t)
        \right]
        \right\}\nonumber\\
        +
        &\frac{1}{80}\sum_k\sum_{j=1}^{15}
        \int_0^t \D{t'}\Gamma_k
        \tr\left\{
        \hat P_{\rm 2Q}\left[\tr_{B}\mathcal{G}\left(\dens_{\hat f_j,B}\right)\right]
        \hat P_{\rm 2Q}
        \tr_{B}\left[
        \hat U(t)
        \mathcal{D}\left[\lop_k(t'),\dens_{\hat f_j,B}\right]
        \hat U^\dag(t)
        \right]
        \right\}, \label{eq:foe2}
    \end{align}
\end{widetext}
\fi
where~$\mathcal{G}$ is now the ideal process, and~$B$ denotes the other degrees of freedom in the system, not included in the ideal gate. We have defined~$\mathcal{U}(\dens) \equiv \hat U(t)\dens\hat U^\dag(t)$ as the unitary part, and~$\dens_{\hat f_j,B}$ is the state of the total system where the qubits performing the gate are set into some Pauli state~$\hat f_j$. For many cases the gate applied on identity simply returns the identity, but for completeness we include it explicitly in Eq.~\eqref{eq:foe2}.

\section{Numerical methods}
\subsection{Time evolution with Magnus expansion}
Solving the time evolution of large time dependent systems requires sophisticated numerical methods. In principle one can always use traditional ODE solvers, such as ZVODE employed by QuTiP~\cite{lambert2024qutip5quantumtoolbox}, but there exists also more sophisticated methods better suited for our needs. The differential equations we solve are of the linear form
\begin{equation}
    \frac{\D{\vec v(t)}}{\D t} = \mat M(t)\vec v(t),
\end{equation}
where~$\vec v$ is some vector (in our case the vectorized density operator), and~$\mat M(t)$ is some time dependent matrix. Formal solution of the equation is
\begin{equation}\label{eq:tdse_sol}
    \vec v(t + \delta t) = e^{\mat U(t+\delta t, t)}\vec v(t),
\end{equation}
where~$\delta t$ is some small time step and the matrix~$\mat U$ is a truncated Magnus series expressed in terms of univariate integrals~\cite{lubich}, and written as
\begin{equation}\label{eq:magnus_expansion}
    \mat U(t+\delta t, t) = \delta t \mat B_0(t) - (\delta t)^2[\mat B_0(t), \mat B_1(t)] + \mathcal{O}[(\delta t)^5],
\end{equation}
with
\begin{equation}
    \mat B_j = \frac{1}{(\delta t)^{j+1}}\int_{-\delta t/2}^{\delta t/2}\D\tau \tau^j \mat M\left(\tau + t + \frac{\delta t}{2}\right).
\end{equation}
In general a single time step using Magnus expansion is computationally much more expensive than one with a conventional ODE solver, due to the need of the calculation of the commutator and the matrix exponential. However, the benefit comes from the fact that Magnus propagation requires considerably less time steps. Moreover, one can actually bypass these heavy calculations by using the Krylov subspace methods described below.

\subsection{Krylov subspace}
Solution in Eq.~\eqref{eq:tdse_sol} is computed efficiently by expressing the matrix~$\mat U$ and the vector~$\vec v$ in an~$m$ dimensional Krylov subspace~$\mathcal{K}_m$ with~$m$ much smaller than the dimension~$d$ of the matrix~$\mat U$. This Krylov subspace is spanned by the vectors
\begin{equation}\label{eq:krylov_basis}
    \{\vec u_1,\, \mat U\vec u_1,\, \mat U^2\vec u_1,\dots,\mat U^{m-1}\vec u_1\},
\end{equation}
where~$\vec u_1\equiv\vec v(t)$. Orthogonalization of this basis gives a~$d\times m$ dimensional unitary matrix
\begin{equation}
    \mat K_m(t,\delta t) = 
    \begin{pmatrix}
        \vec u_1 & \vec u_2 & \vec u_3 & \dots & \vec u_m    
    \end{pmatrix},
\end{equation}
with which the original matrix~$\mat U$ is written as a small~$m\times m$ matrix as
\begin{equation}
    \mat U_m(t+\delta t, t) = \mat K_m^\dag(t,\delta t) \mat U(t+\delta t, t)\mat K_m(t,\delta t),
\end{equation}
and the time evolution of the vector is calculated as
\begin{equation}
    \vec v(t+\delta t) \approx \mat K_m (t,\delta t)e^{\mat U_m(t+\delta t, t)}\mat K_m^\dag (t,\delta t)\vec v(t),
\end{equation}
which only requires the computation of (sparse) matrix vector products and the exponential of a small~$m\times m$ matrix.

If the matrix~$\mat U(t+\delta t, t)$ is Hermitian, one can perform the orthogonalization of the basis in Eq.~\eqref{eq:krylov_basis} with the Lanczos iteration~\cite{saad}, in which case the matrix~$\mat U_m$ becomes tridiagonal,
\begin{equation}\label{eq:tridiag}
    \mat U_m = 
    \begin{pmatrix}
        \alpha_1 & \beta_1 & & & 0\\
        \beta_1 & \alpha_2 & \beta_2 & & \\
         & \beta_3 & \alpha_3 & \ddots & \\
          & & \ddots & \ddots & \beta_{m-1}\\
        0 & & & \beta_{m-1} & \alpha_m
    \end{pmatrix},
\end{equation}
where the matrix elements~$\alpha_j$ and~$\beta_j$, as well as the columns~$\vec u_j$ of the matrix~$\mat K_m$ are obtained from~\cite{Beerwerth2015}
\begin{align}
    \alpha_j &= \vec u_j\cdot \mat U \cdot \vec u_j,\nonumber
    \\
    \beta_j \vec u_{j+1} &= \mat U \cdot \vec u_j - \alpha_j\vec u_j -\beta_{j-1}\vec u_{j-1},\quad\beta_0 = 0.
\end{align}
Note that we here avoid the computation of the commutator in Eq.~\eqref{eq:magnus_expansion}, since we calculate it as a series of sparse matrix-vector products as
\begin{equation}
    \mat U\cdot\vec u = \delta t \mat B_0\cdot\vec u - (\delta t)^2\left[\mat B_0\cdot(\mat B_1\cdot\vec u)
    - \mat B_1\cdot(\mat B_0\cdot\vec u)\right].
\end{equation}
The diagonalization of the tridiagonal matrix in Eq.~\eqref{eq:tridiag} is readily performed with a LAPACK subroutine~\cite{lapack99}, with which the calculation of the matrix exponential can be performed efficiently. The combination of Magnus propagation and Krylov subspace implemented with modern Fortran~\cite{curcic2021toward} results into roughly five times faster simulation time compared to the solver in QuTiP.

\section{Hybridization crosstalk}
\label{app:hybrcrosstalk}
\begin{figure}
    \centering
    \includegraphics[width=1.0\linewidth]{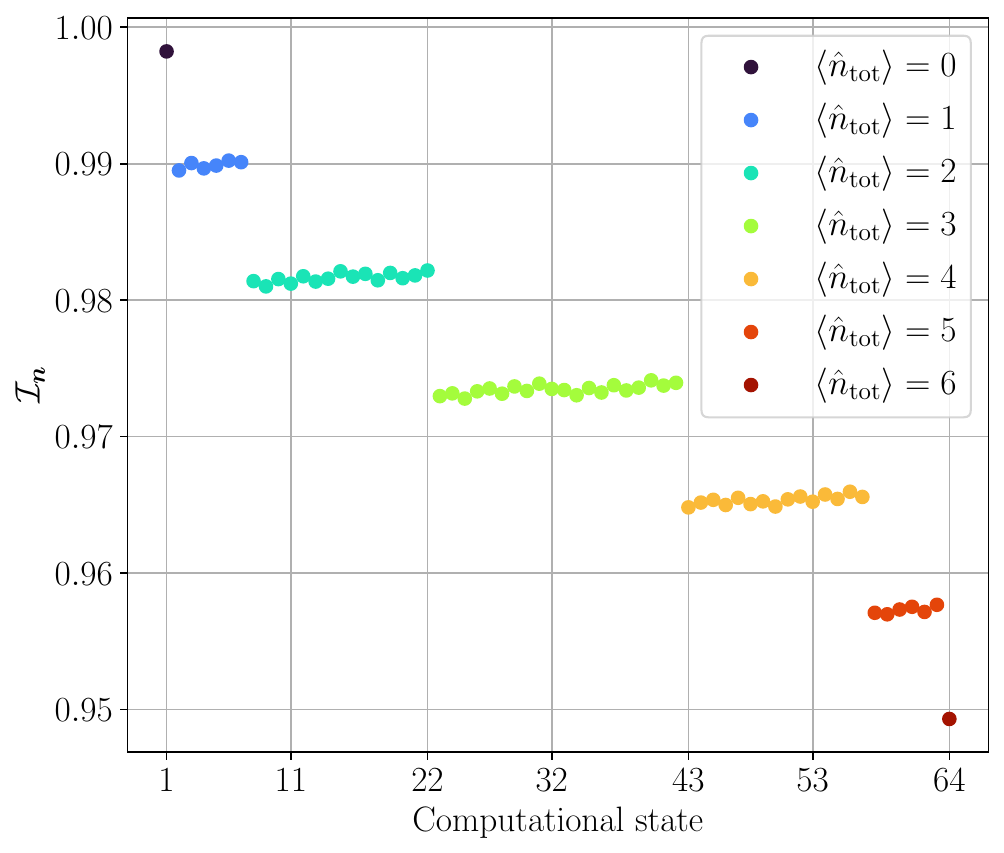}
    \caption{Inverse participation ratios of all 64 computational states shown in Fig.~\ref{fig:compstates}. We have again grouped the states according to the number of excitations in the states.}
    \label{fig:ipr}
\end{figure}

Let us consider a single-qubit gate operation on one of the computational qubits. We here assume without a loss of generality that the qubit $\qub_1$ is the target and we study the influence on the spectator qubit $\qub_2$. Thus, the non-driven system is modelled here with the Hamiltonian in Eq.~\eqref{eq:minimal_hamiltonian_app}. In order to simplify the analytic expressions, we make the rotating-wave approximation for all coupling terms, which is enough to give an intuitive understanding of the physics producing the data in the single-qubit-gate infidelity data shown in Fig.~\ref{fig:sqg}. The single-qubit gate is realized by driving qubit $\qub_1$ with a microwave signal coupled to the local charge operator $\hat q_{\qub_1}=\ii q_{\qub_1}^{\rm zp} (\hat a_{\qub_1}^\dag - \hat a_{\qub_1})$ where $q_{\qub_1}^{\rm zp}$ is the zero-point fluctuation of the charge, see also Eq.~\eqref{eq:sqg_hamiltonian}. 

Since the computational states are eigenstates of the non-driven Hamiltonian, we evaluate the drive-induced transition amplitudes between such eigenstates using fourth-order perturbation theory in all coupling strengths (expressions for the computational states in the local basis are not shown here). Qualitative understanding of the unwanted transitions induced by the drive is obtained by studying the selected matrix elements of the drive operator, which we write in the leading order of the coupling strengths as
\begin{widetext}
    \begin{align}
        \matel{001}{\hat q_{\qub_1}}{000} =& -\ii q_{\qub_1}^{\rm zp} \left[g_{\cent \qub_1}+\frac{g_{\qub_1\tc_1}g_{\cent\tc_1}}{\omega_\cent-\omega_{\tc_1}}\right]\frac{1}{\omega_{\qub_1}-\omega_{\cent}}, \nonumber \\ 
        \matel{010}{\hat q_{\qub_1}}{000} =& -\frac{\ii q_{\qub_1}^{\rm zp}}{\omega_{\qub_2}-\omega_\cent} \left[g_{\cent \qub_1}+\frac{g_{\qub_1\tc_1}g_{\cent\tc_1}}{\omega_{\qub_2}-\omega_{\tc_1}}\right]\left[g_{\cent \qub_2}+\frac{g_{\qub_2\tc_2}g_{\cent\tc_2}}{\omega_{\qub_2}-\omega_{\tc_2}}\right]\frac{1}{\omega_{\qub_1}-\omega_{\qub_2}}, \nonumber \\ 
        \matel{101}{\hat q_{\qub_1}}{100} =&~ \ii q_{\qub_1}^{\rm zp}\left[g_{\cent \qub_1}+\frac{g_{\qub_1\tc_1}g_{\cent\tc_1}}{\omega_\cent-\omega_{\tc_1}}\right]\left(\frac{1}{\omega_{\qub_1}-\omega_{\cent}} - \frac{2}{\omega_{\qub_1}+\alpha_{\qub_1}-\omega_{\cent}}\right), \nonumber \\ 
        \matel{110}{\hat q_{\qub_1}}{100} =& \frac{\ii q_{\qub_1}^{\rm zp}}{\omega_{\qub_2}-\omega_\cent} \left[g_{\cent \qub_1}+\frac{g_{\qub_1\tc_1}g_{\cent\tc_1}}{\omega_{\qub_2}-\omega_{\tc_1}}\right]\left[g_{\cent \qub_2}+\frac{g_{\qub_2\tc_2}g_{\cent\tc_2}}{\omega_{\qub_2}-\omega_{\tc_2}}\right]\left(\frac{1}{\omega_{\qub_1}-\omega_{\qub_2}} - \frac{2}{\omega_{\qub_1}+\alpha_{\qub_1}-\omega_{\qub_2}}\right),\nonumber \\ 
        \matel{200}{\hat q_{\qub_1}}{001} =& -\ii\sqrt{2} q_{\qub_1}^{\rm zp}\left[g_{\cent \qub_1}+\frac{g_{\qub_1\tc_1}g_{\cent\tc_1}}{\omega_\cent-\omega_{\tc_1}}\right]\left(\frac{1}{\omega_{\qub_1}-\omega_{\cent}} - \frac{1}{\omega_{\qub_1}+\alpha_{\qub_1}-\omega_{\cent}}\right), \\ 
        \matel{200}{\hat q_{\qub_1}}{010} =& -\frac{\ii\sqrt{2}q_{\qub_1}^{\rm zp}}{\omega_{\qub_2}-\omega_\cent} \left[g_{\cent \qub_1}+\frac{g_{\qub_1\tc_1}g_{\cent\tc_1}}{\omega_{\qub_2}-\omega_{\tc_1}}\right]\left[g_{\cent \qub_2}+\frac{g_{\qub_2\tc_2}g_{\cent\tc_2}}{\omega_{\qub_2}-\omega_{\tc_2}}\right]\left(\frac{1}{\omega_{\qub_1}-\omega_{\qub_2}} - \frac{1}{\omega_{\qub_1}+\alpha_{\qub_1}-\omega_{\qub_2}}\right). \nonumber
    \end{align}
\end{widetext}
Above, we have assumed that the system idles in a frequency configuration in which the direct couplings between the qubits and the center mode are of the same order of magnitude as the second-order couplings mediated by the couplers. Consequently, we have included terms up to the second order in the couplings $g_{\cent \qub_i}$ and up to the fourth order in the couplings $g_{\ell \tc_i}$. Thus in the following, we interpret that the square-bracketed terms in the expressions for the matrix elements are of first-order in the coupling strengths.

We apply these drive matrix elements when interpreting the infidelity data of single-qubit gates shown in Fig.~\ref{fig:sqg}. 
We observe that due to hybridization, the single-qubit-gate of qubit $q_1$ effectively drives transitions in the center mode with the magnitude that is of the first order in the coupling strengths,
\begin{equation}
    A_{\cent,0\rightarrow 1} = g_{\cent \qub_1}+\frac{g_{\qub_1\tc_1}g_{\cent\tc_1}}{\omega_\cent-\omega_{\tc_1}}.
\end{equation}
Apparently, this magnitude is small at the idling configuration but never exactly zero~\cite{Heunisch2023a}. Despite being small, the matrix elements driving center-mode transitions become large if $\omega_{\qub_1}\approx \omega_\cent$ and $\omega_{\qub_1}\approx \omega_\cent-\alpha_{\qub_1}$. In Fig.~\ref{fig:sqg}, we observe that close to the first resonance there is always an increase in the gate-infidelity. However, even though the matrix element becomes large also close to the latter resonance, the transition $\ket{100}\rightarrow \ket{101}$ is energetically not favourable and, thus, the gate-drive induced center-mode transitions are suppressed even if the target qubit is excited, which is also observed in our numerical data in Fig.~\ref{fig:sqg}.

Interestingly, the gate drive of qubit $\qub_1$ induces transitions also for the qubit $\qub_2$. However, the magnitude of such transitions is of the second-order in the coupling strengths,
\begin{equation}
    A_{\qub_2,0\rightarrow 1} =  
    \left[g_{\cent \qub_1}+\frac{g_{\qub_1\tc_1}g_{\cent\tc_1}}{\omega_{\qub_2}-\omega_{\tc_1}}\right]\left[g_{\cent \qub_2}+\frac{g_{\qub_2\tc_2}g_{\cent\tc_2}}{\omega_{\qub_2}-\omega_{\tc_2}}\right].
\end{equation}
Consequently, compared to the square lattice with single-mode couplers, the coupler-mediated crosstalk is suppressed by our three-mode coupling structure. Furthermore, the crosstalk error is sharply peaked at $\omega_{\qub_1}\approx \omega_{\qub_2}$, as is also observed in Fig.~\ref{fig:sqg}.

Finally, we find that the hybridization-induced crosstalk to the f-state of the target qubit is enhanced close to the resonances $\omega_{\qub_1}\approx \omega_\cent-\alpha_{\qub_1}$ and $\omega_{\qub_1}\approx \omega_{\qub_2}-\alpha_{\qub_1}$. Such crosstalk requires an initial population or simultaneous gate operation of the center mode or the spectating qubit $\qub_2$, respectively. Again, the crosstalk to the spectating qubit is suppressed compared to that of the center-mode, due to the additional modes that mediate and mitigate the coupling.

\section{Inverse participation ratio}\label{app:ipr}

Localization of a quantum state $\ket{\boldsymbol{n}}$ with respect to some fixed complete (and orthonormal) basis $\{\ket{\boldsymbol{m}^0}\}$ is quantified in terms of the inverse participation ratio defined as~\cite{Berke2022,Boerner2023}
\begin{equation}
    \mathcal{I}_{\boldsymbol{n}} = \sum_{\boldsymbol{m}^0}|\braket{\boldsymbol{m}^0}{\boldsymbol{n}}|^4,
\end{equation}
where in our case $\ket{\boldsymbol{n}}=\ket{\boldsymbol{n}_\qub\boldsymbol{n}_\cent}$, where $\boldsymbol{n}_\qub = (n_{\qub_1},n_{\qub_2},n_{\qub_3},n_{\qub_4},n_{\qub_5},n_{\qub_6})$ and $\boldsymbol{n}_\cent = (n_{\cent},n_{\tc_1},n_{\tc_2},n_{\tc_3},n_{\tc_4},n_{\tc_5},n_{\tc_6})$, and the fixed basis is defined by the local eigenstates of the uncoupled transmon system. 
If the state $\ket{\boldsymbol{n}}$ is normalized, the inverse participation ratio of $\mathcal{I}_{\boldsymbol{n}}=1$ indicates that the state is fully localized in the chosen basis. If $\mathcal{I}_{\boldsymbol{n}}<1$, the state is delocalized. 

In Fig.~\ref{fig:ipr}, we show data of the inverse participation ratio $\mathcal{I}_{\boldsymbol{n}}$  computed numerically for the computational states shown in Fig.~\ref{fig:compstates}. The data show that the delocalization increases as the number of excitations within the system increases. Thus, we anticipate that the delocalization-related errors also could be amplified during algorithms that simultaneously excite multiple qubits in the device. However, the inverse participation ratio gives information about the level of delocalization, but not about into which states it is delocalized. Based on Fig.~\ref{fig:compstates}, the strongest hybridization in the computational states is with the local coupler states and, thus,
the delocalization of the computational states to spectating local qubits is small. This potentially mitigates the delocalization-induced errors compared to the square-grid topologies, as was also demonstrated in Fig.~\ref{fig:sqg}.

\end{document}